\documentclass[aps,preprint,showpacs,nofootinbib]{revtex4}
\usepackage{graphicx}
\usepackage{amssymb,amsmath,amsfonts,bm}
\usepackage{color}
\newcommand*{\tcaps}{\setlength{\baselineskip}{3ex}}
\newcommand*{\fs}[1]{#1\!\!\!/}
\def\Ls{{\Lambda^*}}
\def\Tp{{\Theta^+}}
\begin{document}

\title{\boldmath
$\Theta^+$ $formation$ in inclusive $\gamma D\to pK^-X$}

\author{A.\,I.~Titov$^{a,b}$, B. K\"ampfer$^{c,d}$,
S. Dat\'e$^e$, and Y. Ohashi$^e$}

\affiliation{
 $^a$RIKEN, 2-1 Hirosawa,Wako, Saitama 351-098, Japan,\\
 $^b$Bogoliubov Laboratory of Theoretical Physics, JINR,
  Dubna 141980, Russia\\
 $^c$Forschungzentrum Rossendorf, 01314 Dresden, Germany\\
 $^d$ Institut f\"ur Theoretische Physik, TU~Dresden, 01062 Dresden,
 Germany\\
 $^e$Japan Synchrotron Radiation Research Institute, SPring-8,
 1-1-1 Kouto Sayo-cho, Sayo-gun, Hyogo 679-5198, Japan
 }

%%%%%%%%%%%%%%%%%%%% Abstract %%%%%%%%%%%%%%%%%%%%%
\begin{abstract}
 We analyze the possibility to produce an intermediate $\Tp$ via a $KN\to\Tp$
 formation process in $\gamma D\to
 pK^-X$ ($X=nK^+,pK^0$) reactions
 at some specific kinematical conditions, in which a
 $pK^-$ pair is knocked out in the forward direction and its
 invariant mass is close to the mass of $\Ls$
 ($\Ls\equiv\Lambda(1520)$).
 The $\Theta^+$ signal may appear in the $[\gamma D,pK^-]$  missing mass
 distribution.
 The ratio of the signal (cross section at the $\Theta^+$ peak position) to
 the smooth background processes
 varies from 0.7 to 2.5 depending on the
 spin and parity of $\Tp$, and it
 decreases  correspondingly  if the $pK^-$ invariant mass is
 outside of the $\Ls$-resonance region.
 We analyze the recent CLAS search for the $\Tp$ in
 the $\gamma D\to pK^-nK^+$ reaction and show that the conditions of this
 experiment greatly reduce the $\Tp$
 formation process making it difficult to extract a $\Tp$ peak
 from the data.
 \end{abstract}
 \keywords{pentaquark}
 \pacs{14.20.-c, 13.75.Jz, 13.85.Fb}

\maketitle

~\\

\section{Introduction}
 The first evidence for the  pentaquark hadron  $\Theta^+$, discovered by the
 LEPS collaboration at SPring-8 \cite{Nakano03}, was subsequently
 confirmed in some other experiments \cite{OtherPenta}. However,
 many other experiments failed to find the $\Theta^+$ signal
 (for surveys see~\cite{Hicks,Burkert,Danilov}).
 Most of them came from the data analysis of high-statistics
 high-energy experiments. These null results at high energies
 were not so much surprising because it is natural to expect
 a sizable suppression in the production of the more complicated five-quark system
 compared to the conventional three-quark hyperons \cite{THDO}.
 But the state of affairs became dramatic after the recent publication
 of the high statistics results of the CLAS collaboration \cite{JLab-06,DeVita}.
 The first experiment is designed to search for the $\Tp$ signal in
 $\gamma D\to pK^-nK^+$ in direct $\gamma n$ interactions
 at relatively low photon energy, $E_\gamma=1.7-3.5$~GeV.
 The second one aimed to search for the $\Tp$ signal in $\gamma p\to \bar K^0nK^+$
 and $\gamma p\to \bar K^0pK^0$ reactions.
 Within the experimental significance, no $\Tp$ signal was observed.
 Note however, that recently the DIANA collaboration
 confirmed a former result for $\Tp$ production
 in $K^+$ interaction with Xe nuclei \cite{DIANA06}.
 Another positive, but low statistics result on $\Tp$ production
 in $\pi^- p$ interaction was obtained in KEK~\cite{KEK06}.
 Therefore, the situation
 concerning the existence of the  pentaquark state remains  controversial.

 Coming back to the high statistics CLAS experiments, one can conclude
 that
 if the $\Tp$ exists,  then the null result means that we do not understand
 the mechanism of $\Tp$ photoproduction in elementary
 $\gamma N\to \Tp \bar K$ reactions. Indeed, in all
 theoretical studies (for references, see the recent review paper \cite{YNL06})
 the cross section of this reaction is defined by the $K$ and $K^*$ exchange
 dynamics. In the first case, the amplitudes are proportional
 to the product of the $\Tp$-nucleon-kaon coupling constant
 $g_{\Theta NK}$ and the form factor $F(p_\Theta^2,p^2,p^2_K)$,
 where $p_B,p_K$ are the four momenta of the baryon (nucleon or $\Tp$)
 and the kaon, respectively. One of the hadrons is far off-shell.
 If one uses the $\Tp\to NK$ decay width ($\Gamma_\Theta$)
 as an input parameter, then
 the $g_{\Theta NK}$ coupling is fixed, but unfortunately, there are
 no guiding rules for the off-shell form factors which bring some ambiguity
 into the theoretical predictions. For $K^*$ exchange processes the situation
 is even worse. In this case we do not know the $g_{\Theta NK^*}$
 coupling constant (the ambiguity of its estimate is rather
 large \cite{HosakaOset06}) and the ``off-shellness'' in
 the $\Tp$- nucleon-$K^*$ vertex is much greater
 because of the large mass difference between  $K^*$
 and $K$ mesons.
 The CLAS null result for a finite $\Tp$ decay width
 means large off-shell
 suppression of the corresponding amplitudes and small
 $g_{\Theta NK^*}$ coupling constant.

\begin{figure}[th]
{\centering
  \includegraphics[width=.42\textwidth]{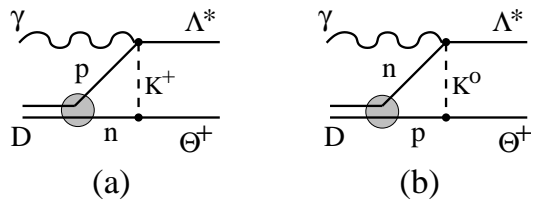}
 \caption{\label{FIG:1}{\small%\tcaps%
 Tree level diagrams for the reaction
 $\gamma D\to \Lambda^*\Theta^+$. The exchange of charged and neutral
 kaons is shown in (a) and (b), respectively.}}}
\end{figure}

 Therefore, the best way to check whether the $\Tp$ exists or not
 is to study the $KN\to\Tp$ fusion reaction  with
 a quasi-free kaon and a nucleon in the initial state.
 In this case the $g_{\Theta NK}$
 coupling is fixed (for given $\Gamma_\Theta$), and there is no
 ambiguity with the off-shell form factor because all hadrons
 are on the mass shell. This situation may be realized in the reaction
 $\gamma D\to \Ls\Tp\to pK^-nK^+$ ($\Ls\equiv\Lambda(1520)$)
 with the  $\Tp$ showing up as a  peak in the $nK^+$ invariant mass
 distribution as shown in Ref.~\cite{TKDO05}.
 There are several conditions which can enhance this
 effect. First, the  $pK^-$ invariant mass must be close
 to the mass of $\Ls$. In this case, the total amplitude is the
 coherent sum of two amplitudes with charged and neutral
 kaon exchange shown in Fig.~\ref{FIG:1}.
 The dominance of the  $K^*$ meson exchange
 in $\Ls$ photoproduction \cite{TKDO05,Barber1980,Sibirtsev05}
 results in a constructive interference between the two amplitudes
 which enhances the $\Tp$  signal.

 Second,
 the deuteron wave function greatly suppresses the processes
 with a fast moving recoil nucleon, therefore, the experiment
 must be able to measure an extremely slowly  moving recoil
 (spectator) nucleon which participates in the $KN\to \Tp\to KN$
 reaction. %(neutron for the CLAS experiment).

 And third, the $pK^-$ pair must be
 knocked out in the forward direction. In this case, the momentum
 of the recoil kaon is small, and it can merge with the
 slowly moving spectator nucleon to produce a $\Tp$.

 The CLAS experiment~\cite{JLab-06} to search for $\Tp$ was designed to study the direct
 $\gamma n\to \Tp K^-\to nK^+K^-$ reaction and,  in
 principle, it does not satisfy the above conditions. Thus,  the $\Tp$
 and the outgoing neutron have finite momenta, and, therefore, the
 experiment has a neutron momentum cut of $p_n>0.2$~GeV/c.
 In order to reduce the contribution of $K^-$ mesons
 coming from $\Ls$ excitation
 the data analysis makes  a cut on the $\Ls$ mass, i.e. the $pK^-$ invariant mass
 is outside the $\Ls$ mass.
 It has cuts for the kaon momenta, $p_K>0.25$~(GeV/c),
 and cuts for the angles
 for positive and negative particles,
  $\theta_+>9$ and $\theta_->15$ degrees, respectively.
 All these  experimental conditions
 (the $pK^-$invariant mass, momenta and the angle cuts)
 while being quite acceptable
 for studying the $\gamma n\to \Tp K^-$ reaction
 result in a large suppression of the $K+N\to\Tp$ formation
 process in the $\gamma D\to pK^-nK^+$ reaction and  reduce the
 ratio of $\Tp$ resonance contribution (signal)
 to background (noise) - S/N.

 In order to avoid the obvious difficulty in measuring the
 slowly moving recoil nucleon one has to  analyze the
 $[\gamma D,pK^-]$ missing mass distribution~\cite{LEPS-prel}. In this
 case,
 all momenta, allowed by the conservation laws participate
 in the process
 and, of course, the dominant contribution would come from slowly
 moving nucleons. As a result, the total cross
 section strongly increases.
 Unfortunately,  in this case the background processes increase
 roughly by a factor of two compared to the exclusive
 $\gamma D\to pK^-nK^+$ reaction,
 because both the $nK^+$ and $pK^0$ final states
 now contribute.
 Nevertheless,
 even under this circumstance such experimental conditions can
 give a better
 chance to see the $\Tp$ signal, in case it exists.
 %Such kind of inclusive measurement is performing now by LEPS collaboration
 %at SPring-8~\cite{LEPS-prel}.

 The aim of the present paper is to extend the results
 of Ref.~\cite{TKDO05} for the
 inclusive reaction  $\gamma D\to pK^-X$, where $X=nK^+,pK^0$,
 towards finding favorable kinematical conditions for
 a manifestation of the $\Tp$ signal. We are going to show that
 this signal is independent
 of the mechanism of the elementary $\gamma N \to\Tp\bar K$
 reaction if the
 $pK^-$ pair is produced in the forward hemisphere.

 Our paper is organized as follows.
 In Sec.~II we consider the kinematics of a $2\to 4$ reaction and
 define
 the observables. In Sec.~III we briefly discuss the elementary
 $\gamma N\to NK\bar K$ reaction which will be used later on
 for estimating the resonant effect and background.
 Sec.~IV is devoted to a description of the associated
 $\Tp\Ls$ photoproduction in $\gamma D$ interactions, where
 we discuss the most favorable kinematics for the coherent effect and the
 dependence of the cross section on $\Tp$ spin and parity.
 In Sec.~V we discuss two dominant components
 of the non-resonant background: spectator and rescattering channels.
 In Sec.~VI we present our main results and give a comparison of a
 possible $\Tp$ signal for the inclusive reaction  $\gamma D\to pK^-X$
 with favorable  kinematics and the exclusive reaction
 $\gamma D\to pK^-nK^+$ under the CLAS conditions. We show that in latter
 case the $\Tp$ signal is weak due to the experimental conditions.
 The summary is given in Sec.~VII.
 In Appendices~A and B we show some details for the
 kinematics considered
 and the amplitudes of the elementary rescattering processes.

\section{Kinematics}

\begin{figure}[t]
{\centering
 \includegraphics[width=.25\textwidth]{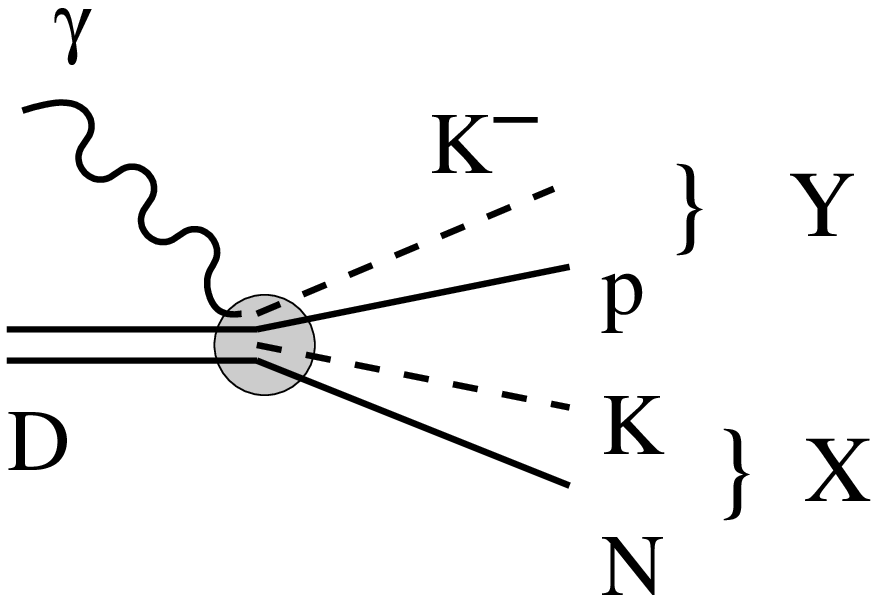}
 \caption{\label{FIG:2}{\small%\tcaps%
 Reaction $\gamma D\to pNK^-K$.}}}
\end{figure}

The differential cross section of the reaction $\gamma D\to
pK^-X$, where $X=nK^+$ or $pK^0$, shown in Fig.~{\ref{FIG:2}},
reads
\begin{equation}
 \frac{d\sigma}{d[...]}\equiv
 \frac{d\sigma}{dM_X dM_Y d\Omega\,d\Omega_{X}\,d\Omega_{Y}}
 =\frac{1}{64\pi^2s_D}\frac{p_f}{p_i}
 \frac{1}{6}\sum\limits_{NK,\lambda,m_D,m_p,m_N}
 |T_{fi}^{NK}|^2\,
 \frac{{\tilde{q}}\,}
 {16\pi^3}\,
 \frac{{\tilde{\bar q}}}
 {16\pi^3}.
  \label{E1}
\end{equation}
Hereafter, we use the following notations: $X$ is the $NK$ pair
with mass $M_X$, $Y$ is the $pK^-$ pair with mass $M_Y$, $p_f$ is
the absolute value of the three-momentum of $Y$  in the $\gamma D$
center of mass system (c.m.s.), $p_i=|{\bf k}|$ is the absolute
value of the photon momentum in the c.m.s., $s_D$ denotes the
square of the total energy in this system, $\tilde{q}$ and
$\tilde{\bar q}$ stand for the absolute values of the $K$ and
$K^-$ mesons momenta in the rest frames of the $X$ and $Y$
systems, respectively. The indices $m_D,m_p,m_N$ correspond to the
spin projections of the deuteron, outgoing proton and nucleon,
respectively, $\lambda$ is the photon helicity; $\Omega_X$ and
$\Omega_Y$ are the solid angles of the directions of flight of $K$
and $K^-$ mesons in the rest frames of the $X$, and $Y$ systems,
respectively; $\Omega$ is the solid angle of the $Y$ system in the
c.m.s. The quantization axis ${\bf z }$ is chosen along the photon
momentum, and the ${\bf y}$ axis is perpendicular to the
production
 plane of $X$ and $Y$ pairs: ${\bf y}={\bf z}\times {\bf
 p}_Y/|{\bf p}_Y|$, where ${\bf p}_Y$ is the three momentum of the
 $Y$ system in the c.m.s.
 $T_{fi}^{NK}$ represents sum of the amplitudes of the resonant
 ($\gamma D\to \Theta^+\Ls \to pK^-X$), semi-resonant
 ($\gamma D\to \Theta^+ pK^- \to pK^-X$), and non-resonant
 ($\gamma D\to pK^-X$) processes.

 The invariant mass distribution $d\sigma/dM_X$ is defined as a
 6-dimensional integral
\begin{equation}
 \frac{d\sigma}{dM_X}
 = 2\pi
  \int  \frac{d\sigma}{d[...]}
  \,d M_Y\,d\cos\theta\,d\Omega_{X}
  d\Omega_{Y}.
  \label{E2}
\end{equation}
In order to define the four momenta of all particles involved in
the process appearing as arguments of the corresponding elementary
amplitudes, we use the following incoming kinematical variables:
photon 4-momentum (laboratory system): $k_L=(E_L,0,0,E_L)$;
deuteron 4-momentum (laboratory system): $p_D=(M_D,0,0,0)$;
invariant masses $M_X$ and $M_Y$; the polar angle of $pK^-$ pair
photoproduction in the c.m.s. $\theta$; and the solid angles
$\Omega_X$ and $\Omega_Y$.

Using these  variables we now calculate all momenta in the $\gamma
D $ c.m.s. (for details see Appendix~A) and then, transform them
to the laboratory system. That is because the deuteron wave
function is only well defined in the laboratory system.

In our study we analyze the missing mass distribution in the range
$M_{\rm min}<M_X<M_{\rm max}$, where $M_{\rm min}=M_N+M_K$ and
$M_{\rm max}=\sqrt{s_D} - M_N - M_K$ in several selected regions
of the invariant mass $M_Y=M_0\pm 20$~MeV. The $KN\to \Tp$
transition leads to a $\Tp$ signal in the missing mass
distribution. Associated $\Ls\,\Tp$ photoproduction manifests
itself  most clearly for $M_0= M_\Ls$ and $M_X\sim M_\Tp$. The
coherent signal must be suppressed outside of the resonance
position. To analyze this situation we choose $M_0$ at the
resonance position with $M_0= M_\Ls$ and at a larger value
($M_0=1.62$~GeV). We also analyze the sensitivity of the $\Tp$
signal to the $pK^-$ pair photoproduction angle to get a maximum
value for the S/N ratio. This gives the conditions for the range
of integration over $\theta$. Integration over $\Omega_{X(Y)}$ in
Eq.~(\ref{E2}) is performed in all regions.

 \section{Elementary $\bm\gamma N\to N \bar KK $ reaction }

 The mechanism of  $\bar KK$ photoproduction
 in $\gamma N$ interaction
 is quite complicated because many processes can contribute.
 In our consideration we select the channel with an
 intermediate excitation of
 $\Ls$, $\gamma N\to \Ls K\to N \bar KK$, and denote it hereafter
 as the "resonant" channel.
 As we will demonstrate, this process is dominant in the associated
 $\Ls\Tp$ photoproduction at $E_\gamma\sim 2$~GeV.

 We denote all other channels as
 "non-resonant" background. Of course, this notation is rather
 conventional, because the $\bar KK$ pairs can also be produced from the
 virtual vector mesons,  hyperon resonances other than $\Ls$, and so on.
 In this case the notation "resonant" selects just the
 $\Ls$ resonance excitation.
 In this work we do not put emphasis on $\Tp$ photoproduction in
 $\gamma N$ interactions because, at the considered kinematics when
 $pK^-$ is produced in the forward direction with a fast moving proton,
 this channel is strongly suppressed by the deuteron wave function.

 In this section all variables are given
 in the $\gamma N$ c.m.s.

\subsection{Reaction $\gamma N\to\Ls K\to N\bar KK $}

 In this part we follow closely our previous paper \cite{TKDO05} and
 recall the main aspects of our
 considerations for the sake of completeness.
 We assume, that at low photon energies,
 close to  threshold, the amplitude of
 $\Ls$ excitation in the $\gamma N\to NK\bar K$ reaction
 may be described by
 the effective Lagrangian formalism, whereas at high energies,
 the Regge model with the $K^*$ exchange as a leading trajectory can be used.
 The value $E_\gamma=2.3$~GeV is chosen as the matching point between
 these two regimes.

\begin{figure}[ht]
{\centering
 \includegraphics[width=.42\textwidth]{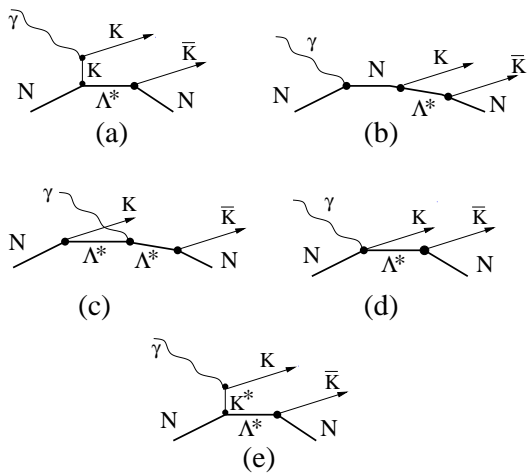}
 \caption{\label{FIG:3}\tcaps%
 Tree level diagrams for the reaction
 $\gamma N\to \Lambda^*{K}\to NK\bar{K}$.}}
\end{figure}

 The tree level diagrams
 for $\gamma N\to \Ls K$ reaction at low energies
 are shown in Fig.~\ref{FIG:3}.
 The diagrams (a), (b), (c) and (d) correspond to the
 $t$, $s$, $u$ exchange amplitudes and the contact term, respectively,
 and are denoted as the Born terms.
 The diagram (e) describes the $t$-channel $K^*$ exchange
 amplitude.
 We neglect the photon interaction within the decay
 vertex and restore the gauge invariance by the proper choice of the
 contact term.

 The amplitudes of the $\gamma p\to\Lambda^*K^+$
 and $\gamma n\to\Lambda^*K^0$ reactions at low energy read
\begin{subequations}
\label{res_amplL}
\begin{eqnarray}
 &&  {A^{\Lambda^*}}_{fi}(\gamma p)=\bar u_{\Lambda^*}^\sigma(p_\Lambda^*)
 \left[
  {{{\cal M}^s}}_{\sigma\mu} +
  {{{\cal M}^t}}_{\sigma\mu} +
  {{{\cal M}^c}}_{\sigma\mu} +
  {{{\cal M}^t}}_{\sigma\mu}(K^*)
  \right]
 u_p(p)\,\varepsilon^\mu~,\\
 && {A^{\Lambda^*}}_{fi}(\gamma n)=\bar u_{\Lambda^*}^\sigma(p_\Lambda^*)
  \left[
 {{{\cal M}^s}}_{\sigma\mu} +
  {{{\cal M}^t}}_{\sigma\mu}(K^*)
  \right]
 u_n(p)\, \varepsilon^\mu~,
\end{eqnarray}
\end{subequations}
\begin{figure}[ht]
 {\centering
 \includegraphics[width=.3\textwidth]{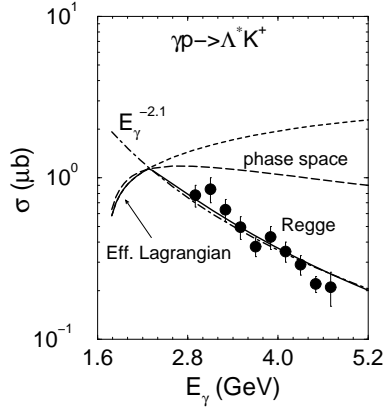}
 \caption{\label{FIG:4}\tcaps%
 The total cross section of the reaction $\gamma p\to \Lambda^*K^+$
 as a function of the photon energy. The experimental data
 are taken from Ref.~\protect\cite{Barber1980}.
 The dot-dashed curve is a fit to this data by
 $\sigma\simeq0.7(\mu$b)$[2.9({\rm GeV})/E_\gamma]^{2.1}$.
 The long-dashed curve
 represents the cross section for a constant amplitude.
 The solid curve corresponds to a solution in the low and high
 energy regimes. The dashed curve describes
 the extrapolation of the effective Lagrangian model to the high
 energy region. See~\protect\cite{TKDO05} for more details.}}
 \end{figure}
\noindent where $u_{\Lambda^*}$, $u_N$ are the $\Ls$ and nucleon
spinors, respectively,  and $\varepsilon^\mu$ is the photon
polarization vector. At high\ energy ($E_\gamma>2.3$~GeV) they are
replaced by the $t$-channel $K^*$ meson exchange amplitude with
Reggeized $K^*$ meson propagator. The explicit form of the
transition operators
 ${\cal M}^{i}_{\sigma\mu}$ as well as the choice of parameters
are given in \cite{TKDO05}.

The total cross section of the reaction $\gamma p\to \Lambda^*K^+$
 as a function of the photon energy from Ref.~\cite{TKDO05}
 together with available
 experimental data \protect\cite{Barber1980} is exhibited
 in Fig.~\ref{FIG:4}.
 Similar results are obtained in  Refs.~\cite{Sibirtsev05,Hosaka0503} using slightly
 different approaches.

\begin{figure}[ht]
{\centering
 \includegraphics[width=.3\textwidth]{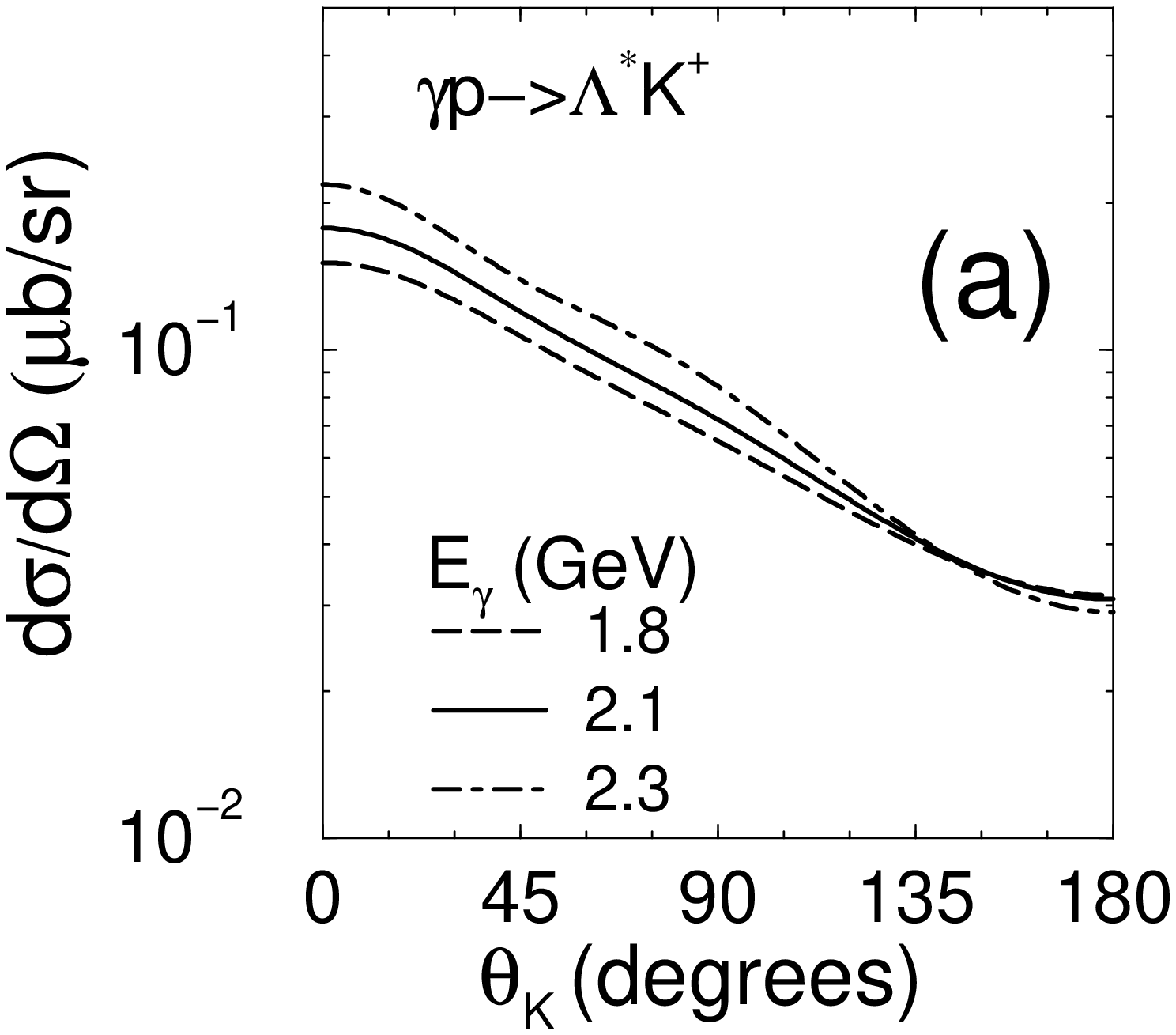}\qquad
 \includegraphics[width=.3\textwidth]{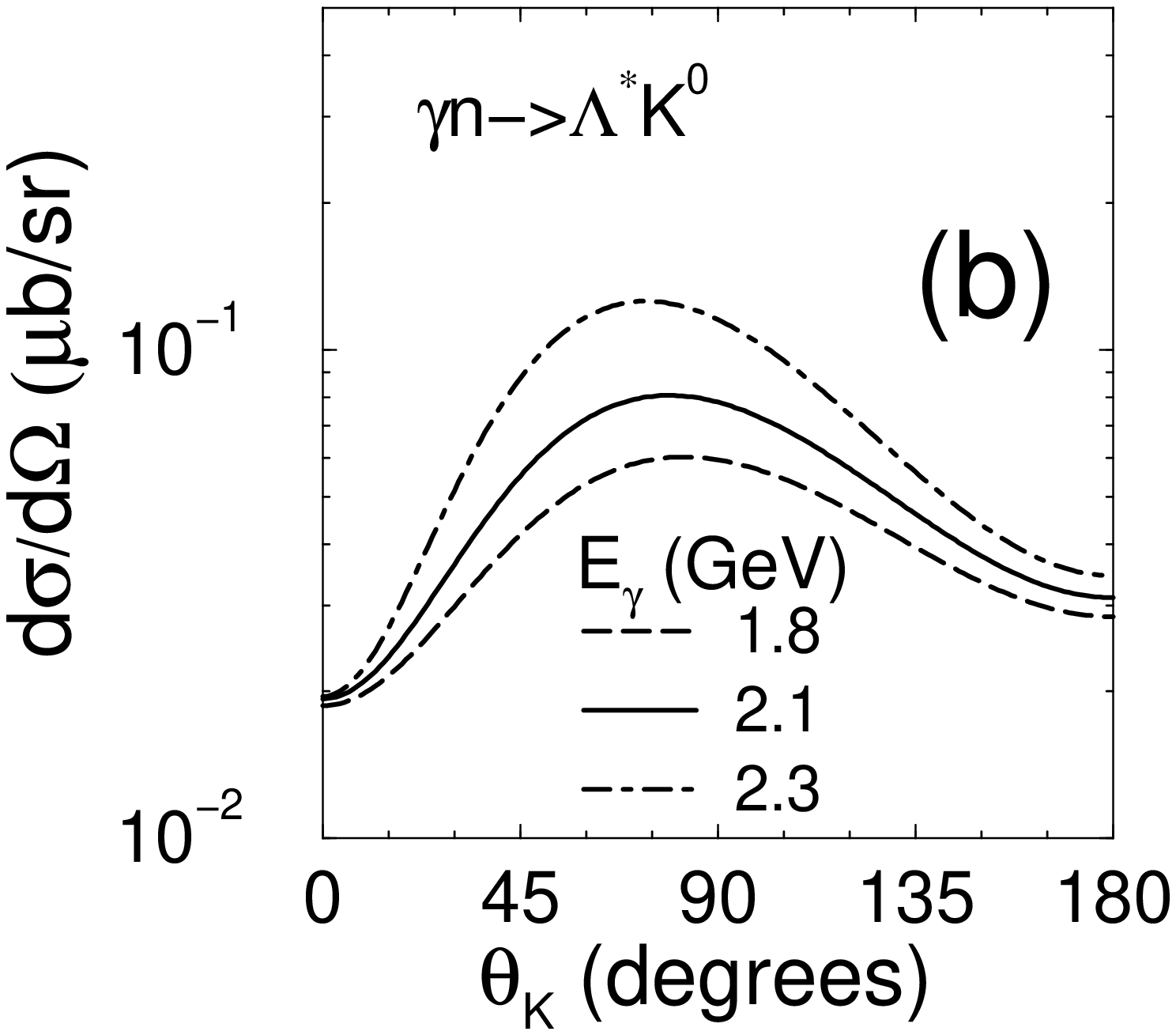}
 \caption{\label{FIG:5}\tcaps%
  The differential cross section of the reactions $\gamma p\to \Lambda^*K^+$
  (a)  and $\gamma n\to \Lambda^*K^0$
  (b)  as a function of the kaon photoproduction angle in $\gamma N$
  c.m.s.  at $E_\gamma=1.8,\,2.1$, and 2.3~GeV. }}
 \end{figure}

\begin{figure}[ht]
{\centering
 \includegraphics[width=.3\textwidth]{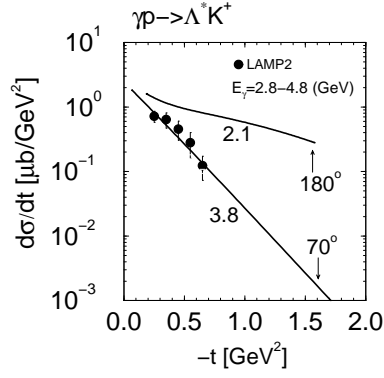}
 \caption{\label{FIG:6}\tcaps%
  The differential cross section for the $\gamma p\to \Lambda^*K^+$
  $E_\gamma=2.1$ and 3.8~GeV. Experimental data from
  Ref.~\protect\cite{Barber1980}.}}
 \end{figure}

   Figure~\ref{FIG:5} shows the differential cross sections for
   $\gamma p\to\Lambda^*K^+$  and $\gamma n\to\Lambda^*K^0$
   as a function of the kaon production angle in the $\gamma N$
   c.m.s.
   at different $E_\gamma$ in the near-threshold region.
   The difference in shape for these two reactions at
   forward photoproduction angles is
   explained by the sizeable contribution of the Born amplitudes
   in the $\gamma p$ reaction.
   In the $\gamma n$ reaction the Born term ($s$-channel) is small,
   and the main contribution comes from the $K^*$ exchange process.
   At backward photoproduction the shapes and the absolute values
   of the cross sections for $\gamma p$ and $\gamma n$
   are similar to each other, but the total cross sections
   for $\gamma p$ is larger. At $E_\gamma=1.8-2.3$~GeV
   it varies from 0.59 to 1.14 $\mu$b as compared with
   0.27 to 1.08 $\mu$b for  $\gamma n$.

   As we will see in  the next section, the dominant
   contribution to the associated $\Ls\Tp$
   photoproduction comes
   from the backward angle of the $K$ photoproduction
   in $\gamma N\to \Ls K$ reaction.
   In Fig.~\ref{FIG:6} we show the differential cross
   section at $E_\gamma=$2.1 and 3.8 GeV together with available
   experimental data~\cite{Barber1980}.
   One can see that for increasing initial photon energy the
   cross section decreases at backward angles for the $K$ photoproduction.
   Therefore, we expect that the threshold region with
   $E_\gamma\leq2.1-2.2$~GeV is most
   favorable for studying associated  $\Ls\Tp$
   photoproduction which reflects the $\Tp$ formation.

   Note that a similar approach for the $\Lambda^*$
   photoproduction based on the effective Lagrangian formalism
   was recently developed by Nam, Hosaka, and Kim \cite{Hosaka0503}.
   The difference to our approach consists in (i) a
   different choice of the form factors and (ii)  a different
   $\Ls NK^*$ coupling constant, which results in an enhancement
    of the Born terms.
   In this case one gets a large constructive interference in the $\gamma p$
   reaction and essentially a destructive interference in the $\gamma n$
   reaction which leads to a strong suppression of the latter one.
   The difference in the different parameter sets
   may be resolved experimentally.
   Finally, let us mention that in our approach
  the total sign of the $\Ls$ photoproduction amplitude follows
  the
  sign of the $K^*$ exchange amplitude. Thus, in the $\gamma p$
  reaction the interference between $K^*$ exchange and Born terms
  is constructive, i.e. their total sign coincide with the sign
  of the $K^*$ exchange amplitude. In the $\gamma n$ reaction the
  $K^*$ exchange is the dominant channel. But SU(3) symmetry
  predicts opposite signs for the $\gamma {K^*}^-K^+$ and
  $\gamma \bar{K^*}^0K^0$ couplings which results in opposite
  signs of the total amplitudes in $\gamma p$ and $\gamma n$
  reactions.

\subsection{Non-resonant  $\gamma N \to N\bar KK $ reactions}

 In Ref.~\cite{TKDO05} we assumed that the dominant contribution to
 the non-resonant $\gamma p\to pK^+K^-$ reaction comes from
 the virtual vector meson
 production ($\gamma p\to p V\to pK^+K^-$)  and intermediate
 $\Lambda(1405)$ excitation ($\gamma p\to \Lambda(1405) K^+\to p
 K^+K^-$). We believe that the vector meson contribution
 is under control because the mechanism of
 real vector meson photoproduction is well known.
\begin{figure}[ht]
{\centering
  \includegraphics[width=.3\textwidth]{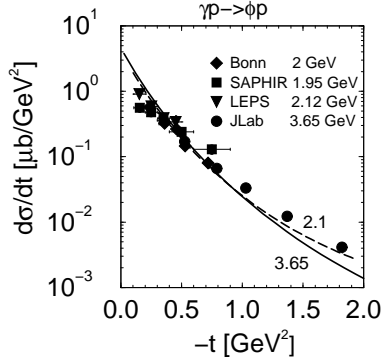}
 \caption{\label{FIG:7}\tcaps%
  The differential cross section of $\phi$ meson
  photoproduction based on the model of Ref.~\protect\cite{TEHN04}.
  Data  from
  Refs.~\protect\cite{Bonn,SAPHIR-PHI,SP8-PHI,JLAB-PHI}.}}
 \end{figure}
 As an example, in Fig.~\ref{FIG:7} we show  the
 differential cross sections
 for $\phi$ meson photoproduction at
 $E_\gamma\sim 2-3$~GeV calculated using the model of Ref.~\cite{TEHN04}
 together with the available experimental data. One can see that the
 description of this reaction is quite reasonable.
 Next,
 the coupling constant of $\phi K^+K^-$ can be extracted
 from the $\phi\to K^+K^-$ decay, and the  $\rho K^+K^-$
 and  $\omega K^+K^-$ couplings can be found from SU(3) symmetry
 relations. Then, the contribution to $K^+K^-$ photoproduction
 from the virtual vector meson excitation may be easily evaluated.
 But at this moment, we have to make two comments.
 First, in the $\gamma N \to N\bar KK $
 reaction the virtual vector mesons are off mass shell and,
 therefore, one has to introduce the
 corresponding form factors \cite{TEHN04}. The form factors, together
 with the
 vector meson propagators, strongly suppress contributions of the
 virtual $\rho$ and $\omega$ mesons leaving only a noticeable contribution
 from the  $\phi$ meson which is almost on-shell because of the small
 decay width of the $\phi$ meson. Second, in the data analysis
 for $\Tp$ photoproduction the contribution of
 the $\phi$ meson can be excluded by making a
 corresponding ``$\phi$-meson cut'' \cite{Nakano03,JLab-06}.
 Nevertheless, we discuss it here in order to fix other sources
 of $K^+K^-$ photoproduction, having in hand only the total
 cross section $\sigma^{K^+K^-}$ of the
 $\gamma p \to p K^+K^-$ reaction~\cite{Erbe:1970cq}.

  The situation with the contribution from
  the $\Lambda(1405)$ is not so transparent.
  At $E_\gamma = 1.8 - 2.5$~GeV, there is some difference
 ($\Delta\sigma^{K^+K^-}$) on the level of $10-30$\%
 between
 the total cross section $\sigma^{K^+K^-}$
 and the total contribution from $\Ls$ and vector meson excitations.
 This difference increases at higher energies, because $\sigma^{K^+K^-}$
 increases with energy, whereas the contribution from $\Ls$ decreases
 with energy and the contribution from the $\phi$ mesons stays constant.
 At low energies, $\Delta\sigma^{K^+K^-}$
 may be identified with the virtual excitation and decay of hyperons other
 than $\Ls$.
 Thus for example,  Oh, Nakayama, and Lee considered
 contributions from
 $\Lambda(1405),\,\Lambda(1116),\,\Sigma(1193)$,
 and $\Sigma(1385)$~\cite{YNL06}, Roberts  included additionally
 contributions
 from  $\Lambda(1600),\,\Lambda(1670),\,\Lambda(1690),\,\Lambda(1800),\,\Lambda(1810),
 \Lambda(1890)$ and
 $\Sigma(1620),\,\Sigma(1660),\,\Sigma(1670),\,\Sigma(1730)
 ,\,\Sigma(1880),\,\Sigma(1940)$~\cite{Roberts04}. In principle, one can also add
 contributions from $\Sigma(1480)$ and $\Sigma(1560)$ hyperons, listed
 in PDG~\cite{PDG}.

 Another source of $\bar KK$ pair
 photoproduction in $\gamma N$ reaction is the so-called Drell
 process \cite{YNL06,Roberts04,Sibirtsev05}
 where the incoming photon
 virtually decays into a $K^+K^-$ pair with subsequent quasi-elastic
 or charge-exchange $KN$ re-scattering.
 Also a $\bar KK$ pair may be produced from the virtual
 decay $\gamma\to KK^*$  with
 a subsequent inelastic $K^*N\to KN$ transition.
 \begin{figure}[ht]
 {\centering
 \includegraphics[width=.4\textwidth]{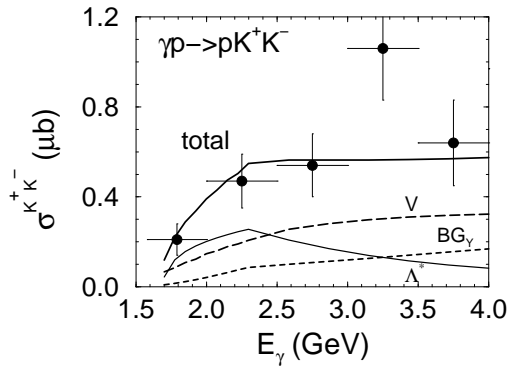}
 \caption{\label{FIG:8}\tcaps%
 The total cross section
 of  the $\gamma p\to p K^+K^-$ reaction
 as a function of the photon energy.
 Thin solid curve is the contribution of $\Ls$ excitation,
 the dashed curve depicts the contribution of the vector mesons decay,
 all other sources denoted as  $BG_Y$ are shown by the
 short-dashed curve.
 The solid curve is the sum of all
 processes. Experimental data
 from Ref.~\protect\cite{Erbe:1970cq}.}}
  \end{figure}
It is quite clear that a consistent description of all the listed
background sources is well beyond the present state-of-the-art
because one needs a fairly large number of poorly known strength
parameters, form factors, phases etc. Moreover, we need a proper
description of the high energy behavior of these processes. On the
other hand, in case of a large number of background sources we can
assume random relative phases between them which leads to
cancellations of the interference terms. Also, as a first
approximation one can choose the incoherent sum of the squares of
the amplitudes to be a constant. This means that the energy
dependence at low energies of this source of $K^+K^-$ pairs is
defined essentially by the phase space factor. Our analysis of the
$\Ls$ photoproduction shows that this approximation works well
(see the solid and long dashed curves in Fig.~\ref{FIG:4} at low
energy). The value of the constant matrix element can be obtained
from a comparison with experimental data for the $\gamma p\to p
K^+K^-$ reaction. In our further
 analysis we parameterize
 the amplitude of the additional contribution $\Delta\sigma^{K^+K^-}$
 (for the sake of a concise notation, we denote it as $BG_Y$)
 by the constant matrix element
 with $|T_{BG_Y}|\equiv T_0=5.95$~GeV$^{-1}$.
 This parameterization, being quite reasonable at low energy
 with $E_\gamma\lesssim 2.3$~GeV, results in a somewhat larger
 rise of the cross section and overestimates
 the data by $20-50\%$ at $E_\gamma=3-6$~GeV.
To fit the data, we multiply $T_0$ by a  correction factor
$C(E_\gamma)$
\begin{eqnarray}\label{CorrFac}
  C^2(E)&=&\theta(E_0-E) +
  \frac{I_C(E_0)}{I_C(E)}\,\left(\frac{E}{E_0}\right)^{1.2}
  \,\theta(E-E_0)~,\nonumber\\
  I_C(E)&=&\frac{1}{s(s-M_N^2)}
  \int\limits_{M_N+M_K}^{\sqrt{s}-M_K}
  \sqrt{\lambda({s,M^2,M_K^2})\,\lambda({M^2,M_N^2,M_K^2})}\,
  \frac{d M}{M}~,\nonumber\\
  s&=&M^2_N + 2M_NE~,
\end{eqnarray}
with the matching point $E_0=2.3$~GeV. In Fig.~\ref{FIG:8} we show
the total cross section of the $\gamma p\to p K^+K^-$ reaction
together with available experimental data~\cite{Erbe:1970cq}.
% The corresponding expressions for the partial contributions are given
% in Appendix B.

 To summarize this section we conclude,  that for the elementary
 $\gamma p\to p K^+K^-$ process which will be used in our analysis
 of the $\gamma D\to pK^-X$ reaction we have selected and described explicitly
 the  $\Ls$ and vector meson ($\phi$ meson) excitation channels.
 The sum of all other possible processes is
 parameterized effectively by a constant matrix element.
 The energy dependence of this channel
 follows the  phase space. At higher energies,
 $E_\gamma = 2.3-3.5$~GeV, this dependence is slightly corrected.
 In case of $\gamma n\to p K^0K^-$ we use the corresponding expressions for
 $\Ls$ photoproduction in the $\gamma n$ reaction,
 keeping the ${BG_Y}$
 contribution the same as in the $\gamma p$ reaction.

 \section{Associated $\bm \Ls\, \Tp $ photoproduction}

Now we turn to the associated $\Ls\,\Tp$ photoproduction off the
deuteron. Basically, our consideration of $\gamma D\to \Tp\Ls $ is
similar to that in Ref.~\cite{TKDO05}, however, we make several
modifications. Therefore, for completeness, we recall the main
aspects of our model to fix the new points. We assume that main
contribution comes from the charge and neutral $K$ meson exchange,
shown in Fig.~\ref{FIG:1}~(a) and (b), respectively, and we do not
discuss the diagrams with direct $\Tp$ photoproduction being
important at backward angles of $pK^-$ pair photoproduction. In
calculating  the $K+N\to\Tp$ vertices we consider the $\Tp$ decay
width as an input parameter, taking
$\Gamma_\Theta=1$~MeV~\cite{SmallWidth}.

 The amplitudes of the associated $\Lambda^*\Theta^+$
photoproduction are expressed through the transition operators of
the "elementary" process $\gamma N\to \Lambda^*K$ as
 \begin{eqnarray}
 A_{(a,b)} = g_{\Theta NK}\int\frac{d^4p}{(2\pi)^4}\,
 \bar u_{\Theta}\gamma_5  \frac{1}{q^2-M_K^2}
 \bar u_{\Lambda^*}^\sigma{\cal M}^{\Lambda^*}_{\sigma\mu}
 \epsilon^\mu
%\nonumber\\
 \,\frac{\fs p + M}{p^2-M^2}\Gamma_D\frac{\fs p' + M}{{p'}^2-M^2}
  U_D~,\label{gamma-D1}
\end{eqnarray}
where the transition operator ${\cal M}$  defines the amplitude of
$\Ls$ photoproduction and uses the sum of transition operators in
Eqs.~(\ref{res_amplL}a) and (\ref{res_amplL}b);
 $\Gamma_D$ and $U_D$ stand for the deuteron $np$
coupling vertex and the deuteron spinor, respectively, $p'=p_D-p$,
and $q$ is the momentum of the exchanged kaon. We  begin our
consideration for the case of $\Tp$ spin-parity $\frac{1}{2}^+$.
Generalization and discussion of our results for another $\Tp$
spin-parity is relegated to the end of this section.

\begin{figure}[th]
{%\centering
 \includegraphics[width=.45\columnwidth]{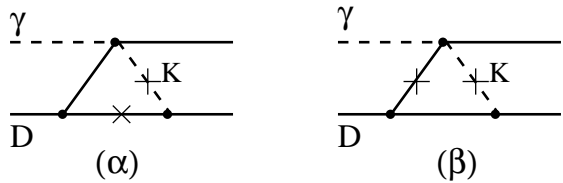}
 \caption{\label{FIG:9}{\small%\tcaps%
  Diagrammatic representation of cutting (indicated  by crosses) the loop
  diagrams.}}}
\end{figure}

Following Refs.~\cite{TKDO05,gammaD} we assume that the dominant
contribution to the loop integrals comes from their pole terms.
The consideration of the regular parts with off-shell kaons needs
incorporation of the corresponding off-shell form factors which
brings an additional ambiguity into the model. Thus, our estimate
may be considered as a lower bound of the coherence effect. The
pole part may be evaluated by summing  all possible cuttings of
the loops, as shown in Fig.~\ref{FIG:9}.
Calculating the imaginary parts we use the following substitutions
 for the propagators of the on-shell particles (Cutkovsky
rules~\cite{Cutkovsky}), shown by crosses in Fig.~\ref{FIG:9},
\begin{eqnarray}
&&\frac{1}{q^2-M_K^2}\to -2\pi i\delta(q^2-M_K^2)~,\nonumber\\
&&\frac{\fs p + M}{p^2-M^2}\to -2\pi i \,(\fs p +
M)\,\delta(p^2-M^2) \label{cut1}
\end{eqnarray}
%\begin{eqnarray}
%-2{\rm Im}\frac{1}{q^2-M_K^2}&=&2\pi\delta(q^2-M_K^2)~,\nonumber\\
%-2{\rm Im}\frac{\fs p + M}{p^2-M^2}&=&2\pi\,(\fs p +
%M)\,\delta(p^2-M^2)~, \label{cut1}
%\end{eqnarray}
and the identity
\begin{eqnarray}
\int{d^4p}\delta(p^2-M^2)=\int\frac{d^3{\bf p}}{2E_p} \label{cut2}
\end{eqnarray}
 with $E_p^2={\bf p}^2+M^2$. We also use the standard representation of the
 product of the deuteron vertex function and the attached nucleon
 propagator through the non-relativistic deuteron wave function
 \begin{eqnarray}
 \Gamma_D\frac{\bar u_{m_1}\bar u_{m_2}}
 {{p'}^2-M_N^2}
 U_{m_D}=\sqrt{2M_D}\,\phi^{m_D}_{m_1m_2}~,
 \label{cut3}
 \end{eqnarray}
 where $p'=p_D-p$,  and $\phi^{m_D}_{m_1m_2}$ is the deuteron wave
 function with the spin projection $m_D$ and
 the nucleon spin projections $m_1$ and $m_2$.
 By using Eqs.~(\ref{cut1}) - (\ref{cut3}),
 one can express the principal parts of the invariant
 amplitudes in Eq.~(\ref{gamma-D1}) as
 \begin{eqnarray}
 \label{gamma-D2}
 A^P&=&\sum\limits_{\xi=\alpha,\beta} A^P(\xi),\nonumber\\
 A^P(\xi)&=&i\frac{\sqrt{2M_D}}{16\pi}
 \sum\limits_{m_1m_2}
 \int\frac{p\,dp}{E_p|{\bf p}_\xi|}
 T^\Ls_{m_1}(\xi)
 \Gamma^{*\Tp}_{m_2}(\xi)\nonumber\\
 &&\qquad\qquad\times\theta(1- a(p,{\bf p}_\xi))
 \,\phi^{m_D}_{m_1m_2}(p,a(p,{\bf p}_\xi)),
 \end{eqnarray}
 where ${\bf p}_\xi$ is the spatial component
 of the corresponding 4-vectors, defined as  $p_\alpha=p_\Tp$ and $p_\beta=p_Y -
 k_\gamma$. Indices $\alpha$ and $\beta$ refer to the left and right cutting
 diagrams in Fig.~\ref{FIG:9}, respectively.
 %%%%%%%%%%%%%%%%%%%%%%%%%%%%%%%%%%%%%%%%%%%%%%%%%%%%%%%%%%%%%
 The function $a(p,{\bf p}_\xi)$ is the cosine of
 the polar angle of the internal nucleon momentum ${\bf p}$ in a deuteron
  when the $z$ axis is along the momentum ${\bf p}_\xi$.
\begin{equation}\label{cosine}
 a(p,{\bf p}_\xi)\equiv \cos\theta_p =
 \frac{M_K^2-M_\xi^2-M_N^2 +2E_\xi E_p}{2|{\bf p}||{\bf p}_\xi|}~,
 \end{equation}
 with $\xi=\alpha,\beta$ and $ M_{\alpha,\beta}^2= p_{\alpha,\beta}^2$.

 The $\Ls$ photoproduction and $\Tp$ decay amplitudes
 read
 \begin{eqnarray}
  \label{gamma-D3}
 T^\Ls_{m_1}(\alpha)&=& \bar u_{\Lambda^*}^\sigma(p_\Lambda^*){\cal
 M}^{\Lambda^*}_{\sigma\mu}\,\epsilon^\mu\, u_{m_1}(p')
 \,\theta({m'}^2),\qquad
 T^\Ls_{m_1}(\beta)= \bar u_{\Lambda^*}^\sigma(p_\Lambda^*){\cal
 M}^{\Lambda^*}_{\sigma\mu}\,\epsilon^\mu\, u_{m_1}(p),
 \nonumber\\
 \Gamma^{\Tp}_{m_2}(\alpha)&=&
 \bar  u_{m_2}(p) \gamma_5 u_{\Theta}(p_\Theta),
 \qquad
 \Gamma^{\Tp}_{m_2}(\beta)=
 \bar  u_{m_2}(p') \gamma_5 u_{\Theta}(p_\Theta) \,\theta({m'}^2), \nonumber\\
 p'&=&p_D -p,\qquad {m'}^2={p'}_0^2 - {\bf p'}^2.
  \end{eqnarray}
 Now we have an additional cut ${m'}^2>0$, compared to Ref.~\cite{TKDO05}
 which suppresses the integrals in Eq.~(\ref{gamma-D2}) and reduces
 the values of the corresponding cross sections.
 The effective deuteron vertex reads
  \begin{eqnarray}
  \phi^{m_D}_{m_1m_2}(p,a)=4\pi \sum\limits_{Lm_Lm_s}
 \langle\frac12 m_1\frac12 m_2|1m_s\rangle
 \langle 1 m_s Lm_L|1m_D\rangle\,i^Lu_L(p)Y_{Lm_L}(\widehat{\bf
  p})~,
 \label{gamma-D4}
 \end{eqnarray}
 where $a$ is the cosine of the polar angle of ${\bf p}$,
 $u_L(p)$ denotes the
 deuteron wave function in momentum space
 \begin{eqnarray}
 u_L(p)=\int  u_L(r)\,j_L(pr)\,rdr,
 \end{eqnarray}
normalized as
 \begin{eqnarray}
 \frac{2}{\pi}\int p^2\left(u^2_0(p) +u^2_2(p)\right)dp=1.
 \label{gamma-D5}
 \end{eqnarray}
 In our calculation we use
 the deuteron wave function derived from the "realistic" Paris
 potential. We checked that the final result does not
 depend on the fine details of the deuteron wave
 function and practically does not depend on the choice
 of the potential.
%In principle, with a good approximation
% one can use the Hulten wave function. In this case Eq.~(\ref{gamma-D4})
% is simplified
%\begin{eqnarray}
% \phi^{m_D}_{m_1m_2}(p,a)=\sqrt{4\pi}
% \langle\frac12 m_1\frac12 m_2|1m_D\rangle u^H(p).
% \label{gamma-D44}
% \end{eqnarray}

Calculating the loop integrals in Eq.~(\ref{gamma-D2}), one has to
be careful with the proper determination of the 3-momentum of
${\bf p}$ which is the argument of the corresponding elementary
amplitudes in the integrals. The azimuthal angle of ${\bf p}$ is
chosen  to be zero because all momenta are in the production
plane. In order to get the internal momentum ${\bf p}$ in the
laboratory system with the $z$ axis along the beam direction, we
make the following transformation
\begin{eqnarray}\label{gamma-D55}
 p_x&\to& p_x\cos\theta_\xi - p_z\sin\theta_\xi\nonumber\\
 p_z&\to& p_x\sin\theta_\xi + p_z\cos\theta_\xi~,
\end{eqnarray}
where $\theta_\xi$ is the polar angle of momentum ${p}_\xi$.

 The differential cross section of the associated $pK^-$ and $NK$
 photoproduction, integrated over the $pK^-$ invariant mass in the
 range $M_Y=M_\Ls\pm 20$~MeV at
 $M_X=M_\Tp$ is related to the differential cross section of the
 associated $\Ls\Tp$ photoproduction as
\begin{equation}\label{gamma-D6}
  \frac{d\sigma^{\gamma D\to pK^-X}}{d\cos\theta
  \,dM_X}\Big|_{M_X=M_\Tp}
  \simeq
  \frac{N}{\pi\Gamma_\Tp}\,
  \frac{d\sigma^{\gamma D\to \Lambda^*\Theta^+}}{d\cos\theta}~,
\end{equation}
where $N\simeq 0.17$ is the integral over the Breit-Wigner $\Ls\to
pK^- $ decay distribution
\begin{equation}\label{gamma-D7}
  N=  B_{pK^-}\, \int\limits_{M_\Ls-\Delta}^{M_\Ls+\Delta}
  \frac{2M_\Ls M_X\Gamma_\Ls dM_X}
  {(M_X^2-M_\Ls^2)^2 + (M_\Ls\Gamma_\Ls)^2}
\end{equation}
with $\Delta=20$~MeV and the branching ratio $B_{pK^-}\simeq
0.45/2$.

 The differential cross section of the coherent $\Lambda^*\Theta^+$
 photoproduction reads
 \begin{eqnarray}
 \frac{d\sigma^{\gamma D\to \Lambda^*\Theta^+}}{d\cos\theta}
 =\frac{1}{32 \pi}\frac{1}{s_D}\frac{p_{f}}{p_{i}}\, |A_{a}+A_{b}|^2~,
 \label{cs-ch}
 \end{eqnarray}
 where $A_a$ and $A_b$ are the amplitudes of the charge and neutral current
 exchange, respectively, depicted in Fig.~\ref{FIG:1}.
 In this equation, averaging and summing
 over the spin projections in the initial
 and the final states are performed. The difference between $A_a$
 and $A_b$ consists of different elementary amplitudes for the
 $\gamma p\to \Ls K^+$ and $\gamma n\to \Ls K^0$ reactions, and
 in an opposite sign of the $\Tp nK^+$ and  $\Tp pK^0$ couplings
 which is a consequence of the zero isospin of $\Tp$.
 The relative sign of the amplitudes of
 $\gamma p\to \Ls K^+$ and $\gamma n\to \Ls K^0$ follows
 the relative sign of the $\gamma \bar {K^0}^*K^0$ and
 $\gamma {K^-}^*K^+$ coupling constants and, according to SU(3) predictions,
 is opposite. Therefore, the sum of the charged and neutral
 $K$ meson exchange diagrams leads to a
 constructive interference between $A_a$ and $A_b$, and
 an enhancement of the
 cross section of the associated $\Ls\Tp$ photoproduction.
\begin{figure}[th]
{%\centering
 \includegraphics[width=.3\columnwidth]{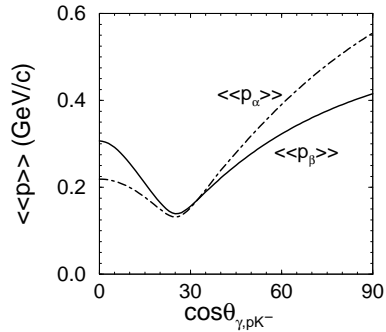}
 \caption{\label{FIG:10}{\small%\tcaps%
   The average momenta $\langle\langle p_\alpha\rangle\rangle$
and $\langle\langle p_\beta\rangle\rangle$ in the loop diagrams
   shown in Figs.~\protect\ref{FIG:9} ($\alpha$) and ($\beta$),
   respectively,
   as a function of the $pK^-$ photoproduction angle.}}}
\end{figure}

In Fig.~\ref{FIG:10} we show the average momenta $\langle\langle
p_\alpha\rangle\rangle$ and $\langle\langle p_\beta\rangle\rangle$
in the loop diagrams
   depicted in Figs.~\ref{FIG:9} ($\alpha$) and ($\beta$),
   respectively, as a function of the $pK^-$
   photoproduction angle in the c.m.s.
   This example corresponds to $E_\gamma=2.1$~GeV and $M_0=1.52$~GeV.
 The definition of this averaging is given as usual,
\begin{equation}\label{Pav}
\langle\langle p \rangle\rangle^2 =\frac{\int dM_X d\Omega_X
d\Omega_Y\, \langle p\rangle^2} {\int dM_X d\Omega_X d\Omega_Y}~,
\end{equation}
with
\begin{equation}\label{Pav_}
  \langle p\rangle =\frac{\int pF(p)dp}{\int F(p)dp}~,
\end{equation}
where $F(p)$ is the integrand in the loop integrals. One can see
that the average momenta have a minimum at $\theta\sim25^0$. Near
this position the corresponding amplitudes have a maximum. At
large angles, mean values of $p$ are large and as a result, the
corresponding amplitudes are very small because of the
exponentially small value of the deuteron wave function at large~$p$.

\begin{figure}[th]
{%\centering
 \includegraphics[width=0.3\columnwidth]{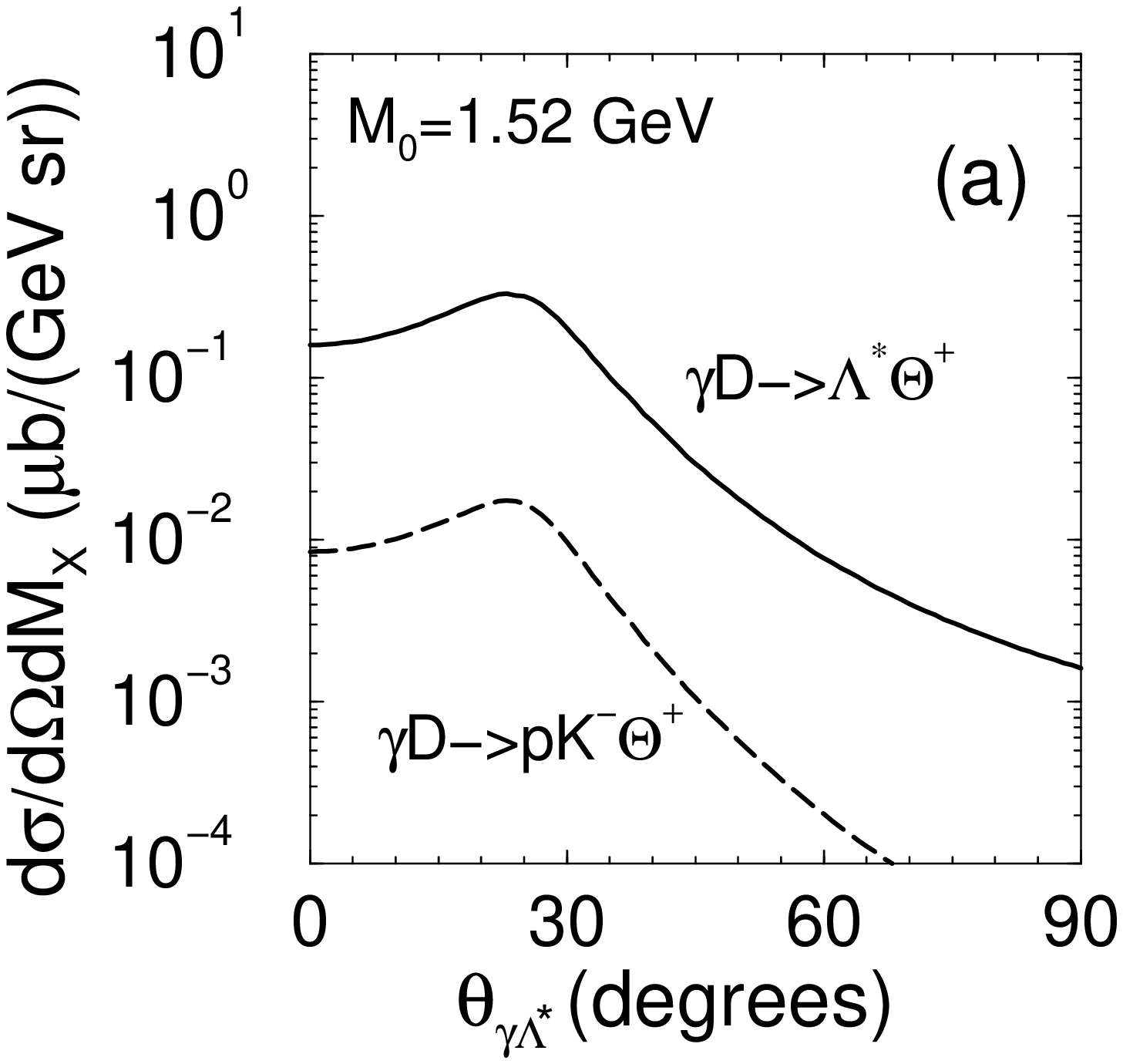}\qquad
 \includegraphics[width=0.3\columnwidth]{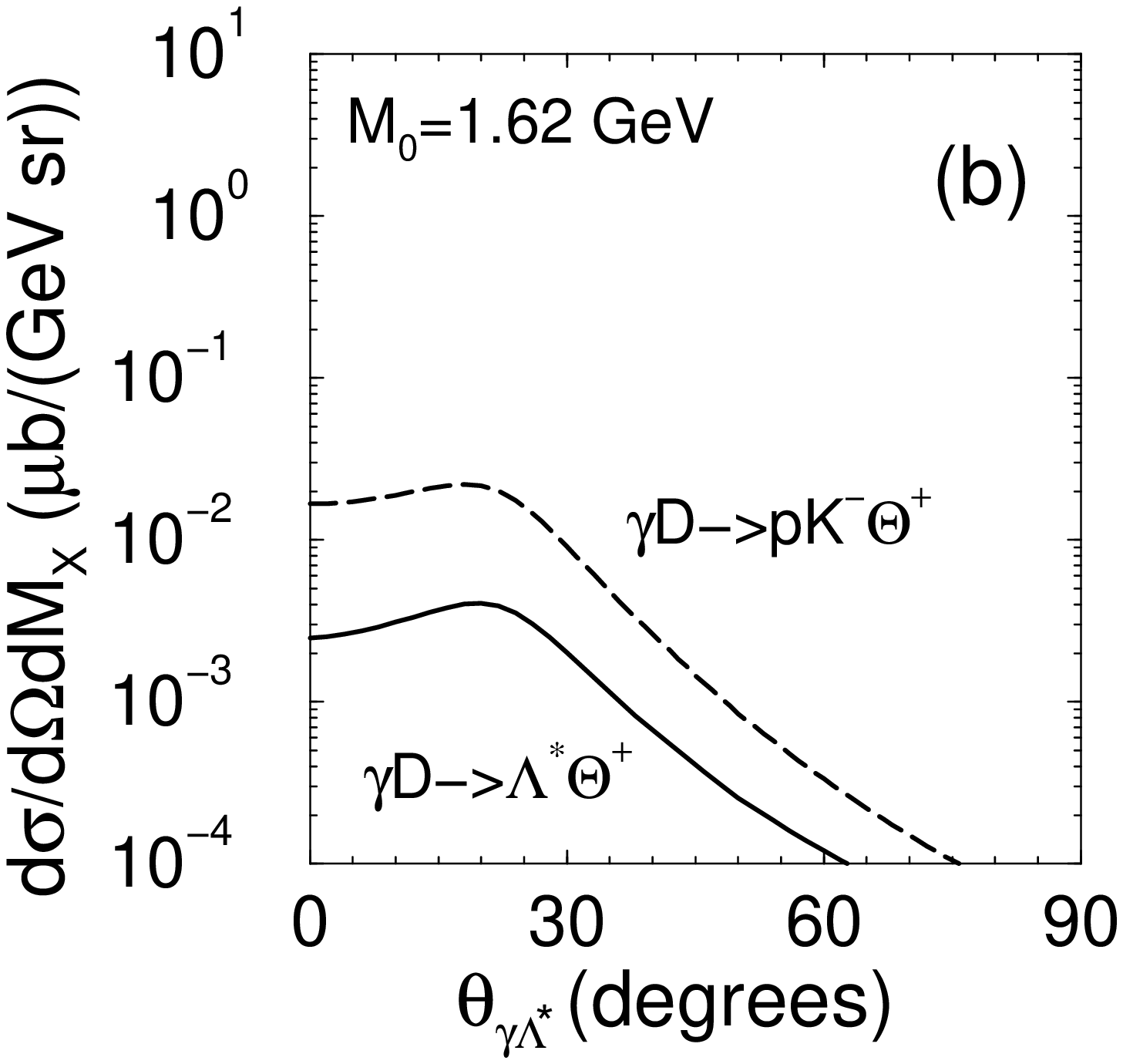}
  \caption{\label{FIG:11}{\small%\tcaps%
  The angular distribution of the
  differential cross section
  $d\sigma/d\Omega dM_{X}$ at
  $M_X=M_\Tp=1.53$~GeV,  $E_\gamma=2.1$~GeV and
  $M_0=M_\Ls=1.52$
  and 1.62 (GeV), shown in (a) and (b), respectively.
  The solid and dashed curves correspond to resonant
  and coherent background contributions.}}}
 \end{figure}

\begin{figure}[th]
{%\centering
  \includegraphics[width=0.45\columnwidth]{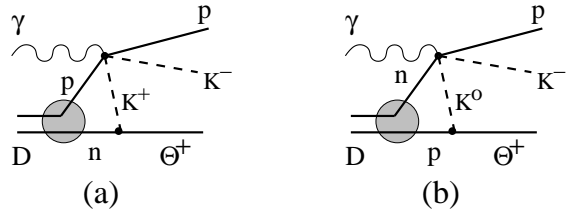}
  \caption{\label{FIG:12}{\small%\tcaps%
  Diagrammatic representation of the associated non-resonant $pK^-\Tp$
  photoproduction.}}}
 \end{figure}

 In Fig.~\ref{FIG:11} we show the angular distribution of the
 differential cross section $d\sigma/d\Omega dM_{X}$ of the
 $\gamma D\to \Theta^+ pK^-\to pK^-X$ reaction at
 $E_\gamma=2.1$~GeV,
 and  the missing mass $[\gamma D,\,pK^-]$, $M_X=M_\Tp=1.53$~GeV,
 and  at $M_0=M_\Ls=1.52$ and 1.62~GeV in (a) and (b),
 respectively.
 The solid curves correspond to the resonance contribution, i.e.
 $\gamma D\to\Ls\Tp \to pK^-X$, while the dashed curves shows
 the contribution from the non-resonant $\gamma N\to pK^-K$ processes
 depicted schematically in Fig.~\protect\ref{FIG:12}.\\
%%%%%%%%%%%%%%%%%%%%%%%%%%%%%%%%%%%%%%%%%%%%%%%%%%%%%%%%%%%%%%%%%
%%%%%%%%%%%%%%%%%%%%%%%%%%%%%%%%%%%%%%%%%%%%%%%%%%%%%%%%%%%%%%%%%
\begin{figure}[th]
{%\centering
  \includegraphics[width=0.3\columnwidth]{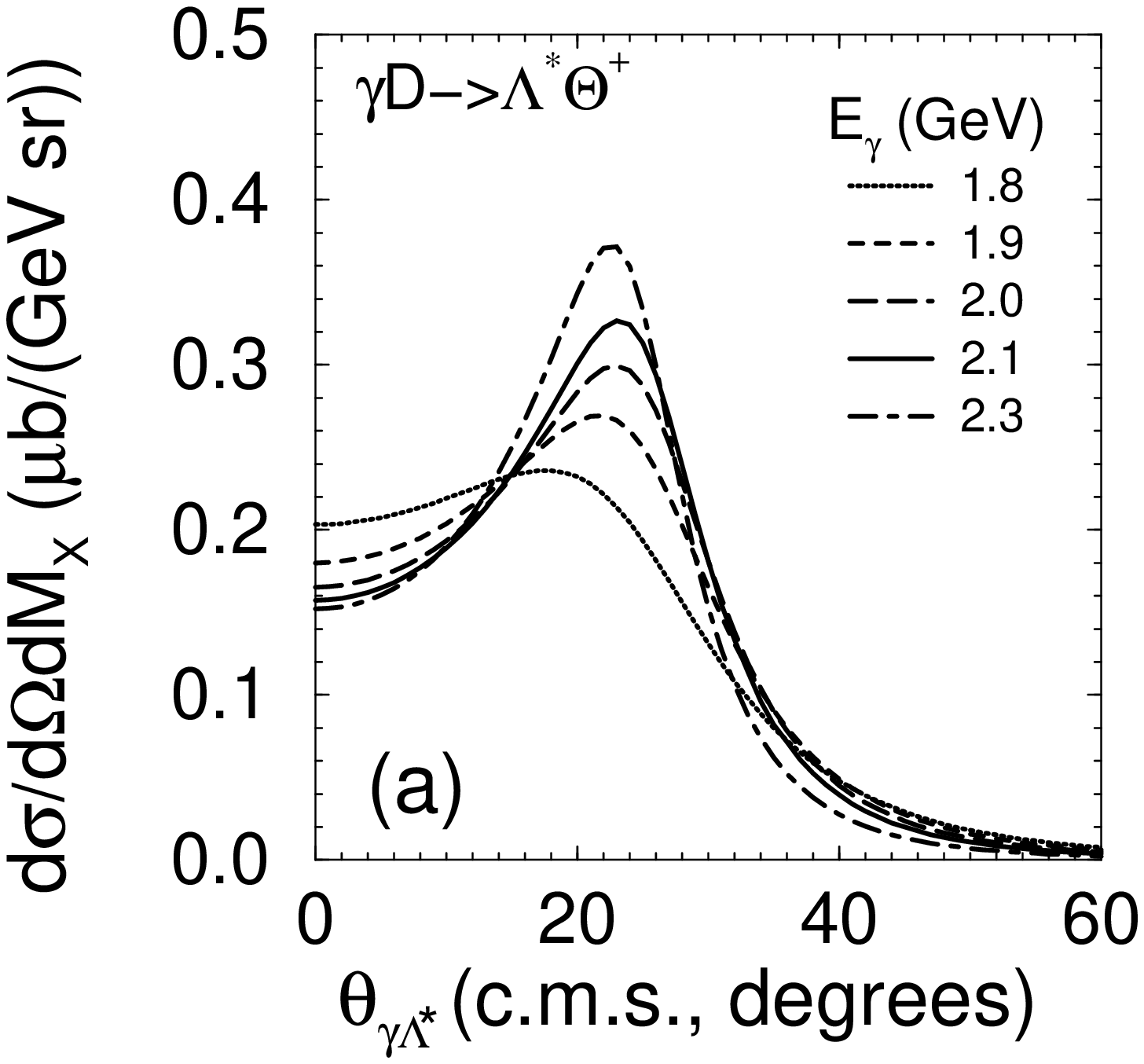}\qquad
  \includegraphics[width=0.3\columnwidth]{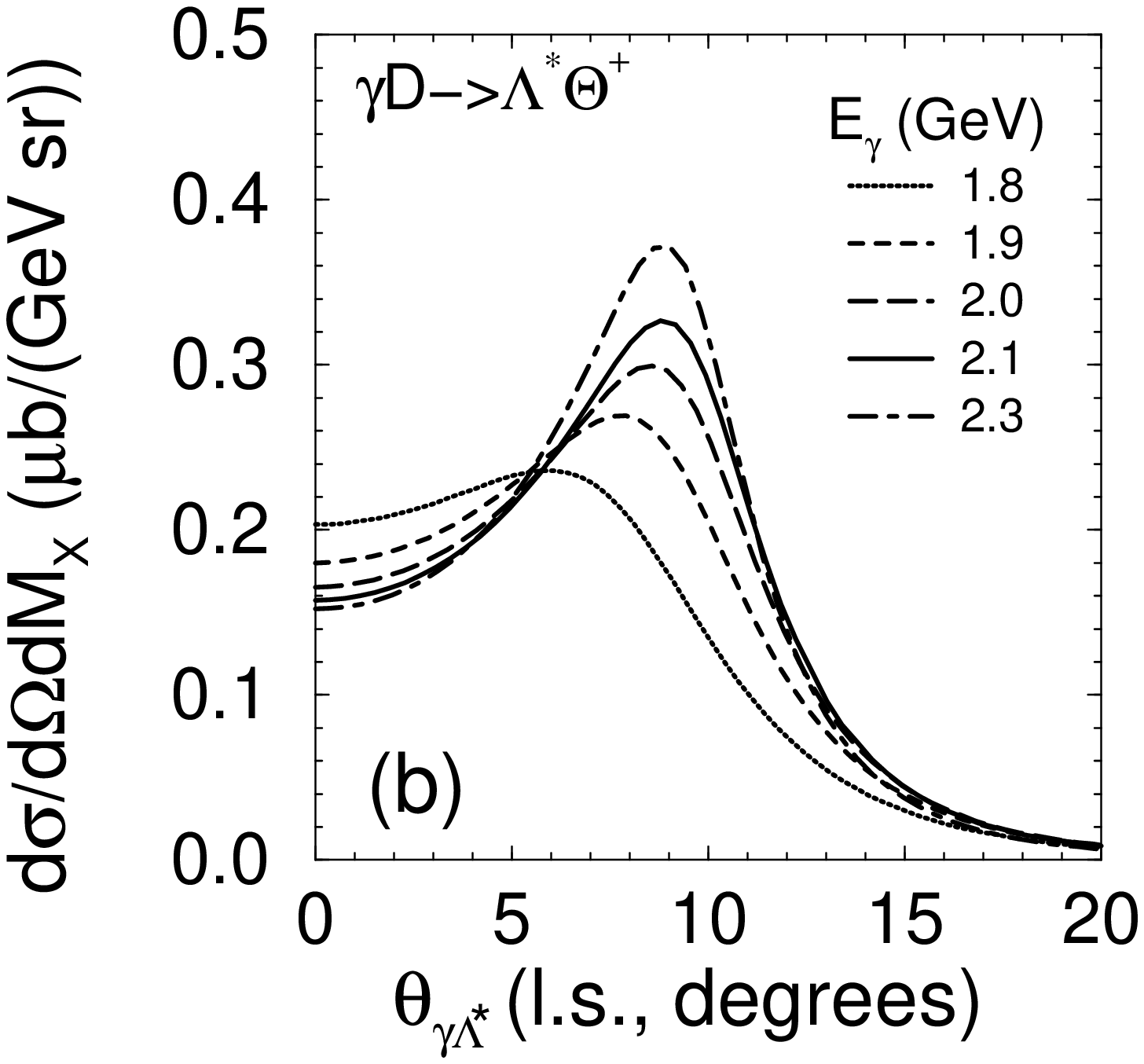}
  \caption{\label{FIG:13}{\small%\tcaps%
 The same as in Fig.~\protect\ref{FIG:11},
 but for different $E_\gamma$.
 (a) and (b) correspond to the dependence on the
 $\Ls$ photoproduction angle in the c.m.s. and
 the laboratory frame, respectively.}}}
 \end{figure}
One can see, that in the $\Ls$ region, where $M_0=M_\Ls$, the
resonant cross section is about one order of magnitude larger than
the contribution of the non-resonant channels discussed in the
previous section. Outside the resonance position, say for
$M_0=M_\Ls+100$~MeV, the result is opposite, namely, the resonant
contribution is strongly suppressed because of a small  $\Ls$
total decay width, and the processes with non-resonant $\gamma
N\to pK^-K$ transitions become dominant.
%%%%%%%%%%%%%%%%%%%%%%%%%%%%%%%%%%%%%%%%%%%%%%%%%%%%%%%%%%%%%%%%%
%%%%%%%%%%%%%%%%%%%%%%%%%%%%%%%%%%%%%%%%%%%%%%%%%%%%%%%%%%%%%%%%%
In Fig.~\ref{FIG:13} (a) and (b) we show the resonant cross
section as a function of the $\Ls$ photoproduction angle in the
c.m.s. and the laboratory system, respectively, for several values
of the photon energy. The value of the cross section at maximum
and the position of the maximum depends on the energy.
%The position of the
%maximum of the differential cross section as a function of the
%photon energy is shown in Fig.~\ref{fig:13-2}.
One can see that
the $\Tp$ formation in associated $\Ls\Tp$ photoproduction is
hardly measurable if the detector acceptance does not allow measuring
the $pK^-$ pairs at small angles $\theta_{\rm lab} \le 10^0$.
%%%%%%%%%%%%%%%%%%%%%%%%%%%%%%%%%%%%%%%%%%%%%%%%%%%%%%%%%%%%%%%%

Finally, let us discuss the dependence of the associated $\Ls\Tp$
photoproduction on the spin and parity of the $\Tp$.  The case of
$J^P_\Theta=3/2^-$ is especially attractive because the small
$\Tp$ decay width~\cite{SmallWidth} has a natural explanation for
this assignment of the $\Tp$ spin and
parity~\cite{Theta3half0,Theta3half1,Theta3half2,Theta3half3}.

The effective Lagrangians of the $\Tp N K$ interactions  are
expressed usually in the following form~\cite{Theta3half2}
\begin{eqnarray}
 {\cal L}^{\frac12^\pm}_{\Theta NK}
  &=&  {g^{\frac12^\pm}_{\Theta NK}}
  \bar \Theta \Gamma^{\pm} K N  + \text{h.c.}~,
  \label{RES2}\\
  \nonumber\\
 {\cal L}^{\frac32^\pm}_{\Theta NK}
  &=&  \frac{g^{\frac12^\pm}_{\Theta NK}}{M_\Theta}
  \bar \Theta^\alpha \Gamma^{\mp} ({\partial_\alpha K})\, N  + \text{h.c.}~,
  \label{RES3}
\end{eqnarray}
where $\Theta, N$ and $K$ are the $\Tp$, nucleon and kaon fields,
$\Gamma^+=\gamma_5$, and  $\Gamma^-=1$. For the fixed $\Tp\to NK$
decay width the coupling constant $g_{\Theta NK}$ depends on the
spin and parity  of $\Tp$ as
\begin{eqnarray}
 \label{RES4}
 |g_{\Theta NK}^{\frac12^\pm}|^2 &=&
 \frac{4\pi\Gamma_\Theta}{p_F}
 \frac{M_\Theta^2}{(M_\Theta\mp M_N)^2 -M_K^2}~,
\nonumber\\
 \label{RES5}
 |g_{\Theta NK}^{\frac32^\pm}|^2 &=& \frac{48\pi\Gamma_\Theta}{p_F}
 \frac{M_\Theta^6}{
 \lambda(M_\Theta^2,M_N^2,M_K^2) [(M_\Theta\pm M_N)^2 -M_K^2]}~,
\end{eqnarray}
where $p_F= \sqrt{\lambda(M_\Theta^2,M_N^2,M_K^2)}/2M_\Theta$ is
the $\Tp$ decay momentum. These equations result in the following
relation
\begin{equation}\label{RES6}
 |g_{\Theta NK}^{\frac12^-}|:
 |g_{\Theta NK}^{\frac12^+}|:
 |g_{\Theta NK}^{\frac32^+}|:
 |g_{\Theta NK}^{\frac32^-}|=0.134:1:1.39:10.21~,
\end{equation}
where, for example for $\Gamma_\Theta=1$~MeV,  $|g_{\Theta
NK}^{\frac12^+}|=1.04$. Using this estimate one can expect naively
that in case of $J^P=\frac32^-$ ($\frac12^-$) the coherent
$\Ls\Tp$ photoproduction would be enhanced (suppressed) roughly by
two orders of magnitude compared to the case of $J^P=1/2^+$. But
the real situation is far from this expectation. The matrix
elements defining the  $\Tp$ formation are proportional to the
products
\begin{equation}\label{RES7}
g_{\Tp NK}\times t_{m_\Theta m_N}
\end{equation}
with
\begin{eqnarray}
 t_{m_\Theta m_N}^{\frac12^\pm}&=&
 {{\bar{u}}_\Theta}{}_{{m_\Theta}}\,\Gamma^\pm\,
 u_{m_N}~,\nonumber\\
 t_{m_\Theta m_N}^{\frac32^\pm}&=& {{\bar
 {u}}_\Theta^\alpha}{}_{{m_\Theta}}\,\Gamma^\mp\, q_\alpha\,
 u_{m_N}~,\label{RES8}
\end{eqnarray}
where $q$ is the kaon momentum, $m_\Theta$ and $m_N$ denote the
spin projections of $\Tp$ and nucleon, respectively. As a result,
the large (small) value of $|g_{\Theta NK}|$ is compensated by the
corresponding small (large) value of $t_{m_\Theta,m_N}$. For a
qualitative estimate of such a compensation let us consider the
combination
\begin{equation}\label{RES9}
  |A|^2=\sum\limits_{m_\Theta m_N}|g_{\Theta NK}\,t_{m_\Theta,m_N}|^2~,
\end{equation}
where the nucleon may be off-shell and express $|A|^2$ through the
$\Tp$ decay width
\begin{eqnarray}
 |A^{\frac12^\pm}|^2&=&8\pi M_\Theta^2\frac{\Gamma_\Theta}{p_F}
 \frac{(M_\Theta\mp \bar M_N)^2 -M_K^2}{(M_\Theta\mp M_N)^2 -
 M_K^2}~,
 \nonumber\\
 |A^{\frac32^\pm}|^2&=&16\pi M_\Theta^2\frac{\Gamma_\Theta}{p_F}
 \frac{\lambda(M_\Theta^2,\bar M_N^2,M_K^2)}{\lambda(M_\Theta^2, M_N^2,M_K^2)}
 \frac{(M_\Theta\pm \bar M_N)^2 -M_K^2}{(M_\Theta\pm M_N)^2 - M_K^2}~,
 \label{RES10}
\end{eqnarray}
 where $\bar M_N^2=p^2>0$ is the square of the nucleon momentum in the $\Tp
 NK$ vertex.
 From these equations one can conclude that in case of an on-shell
 nucleon with $\bar M_N^2=M_N^2$:
 (i) $|A|^2$ does not depend on parity, and (ii)
 $|A^{\frac32}|^2=2|A^{\frac12}|^2$. The latter
 is the consequence of the spin
 factor $2J+1$ in  the expression for the decay width. The dependence
 on parity arises only for off-shell nucleons.
 $|A|^2$ increases (decreases)
 for $J^P=1/2^+$ and $3/2^\pm$ ( $1/2^-$)
 at the off-shell
 region with $\bar M_N^2<M_N^2$.
 The increase for $J^P=3/2^+$ and $3/2^-$ is different, because in
 the former case $|A|^2$ is defined by the interplay of suppression
 and enhancement factors. On average, the ratio of
 $|A^\frac32|^2/|A^\frac12|^2$ would be slightly larger than 2.
 \begin{figure}[th]
 {%\centering
  \includegraphics[width=0.3\columnwidth]{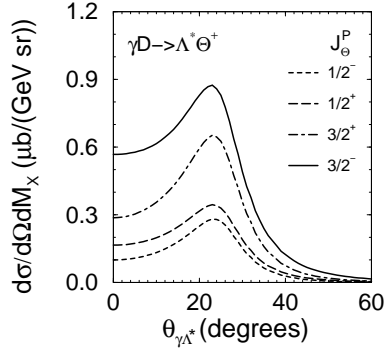}
  \caption{\label{FIG:14}{\small%\tcaps%
  The same as in Fig.~\protect\ref{FIG:13}~(a), but for different
  $\Tp$ spin and parities. $E_\gamma=2.1$~GeV and $M_0=1.52$~ GeV.
  }}}
  \end{figure}

 In some sense, the dependence of the amplitude of the associated
 $\Ls\Tp$ photoproduction on spin and parity of the $\Tp$ is rather
 similar to that of $A^{J^P}$ in our example.
  The diagrams with the on-shell hadrons  in the $\Tp$ formation
  vertex
 (see Fig.~\ref{FIG:9} ($\alpha$)) do not depend on $\Tp$ parity, and
  for $J=3/2$ their contribution is roughly two times greater than
  for $J=1/2$. The amplitudes with the off-shell nucleon in
   the $\Tp$ formation vertex
  (Fig.~\ref{FIG:9} ($\beta$)) are
  enhanced for $J^P=1/2^+,3/2^\pm$ and suppressed for $J^P=1/2^-$.
  But this off-shell modification is not so large, because
  the contribution of the nucleons which are far off-shell
  are suppressed by the deuteron wave function.

  In Fig.~\ref{FIG:14} we show
  the angular distribution of the
  differential cross section
  $d\sigma/d\Omega dM_{X}$ at
  $M_X=M_\Tp=1.53$~GeV, and  $E_\gamma=2.1$~GeV,
  $M_0=M_\Ls=1.52$ and different $\Tp$ spin and parities,
  $J^P=\frac12^\mp$ and $\frac32^\pm$.
  The ratio of the cross section at their maximum position
  for different $J^P$  reads
 \begin{equation}\label{RES11}
 \frac12^-:\frac12^+:\frac32^+:\frac32^-\simeq
  0.81:1:1.87:2.53~.
 \end{equation}
 This result is in agreement with our qualitative analysis, namely,
 the cross section for $J=3/2$  on average is  2.4 times greater
 than for $J=1/2$. For $J^P=1/2^+$ and $3/2^-$ the cross sections
 are enhanced compared to the cases of $J^P=1/2^- $ and $3/2^+$,
 respectively.

%%%%%%%%%%%%%%%%%%%%%%%%%%%%%%%%%%%%%%%%%%%%%%%%%%%%%%%%%%%%%%%%%
 Now two questions arise.
 First, whether the associated $\Ls\,\Tp$
photoproduction may be seen against other non-resonant processes
in the resonance region with  $M_0=M_\Ls$, and second, whether
this signal is suppressed outside the resonance region and why? To
answer these questions we have to analyze the background
processes.

\section{Non-resonant background}

\subsection{Spectator channels}

\begin{figure}[th]
 {%\centering
  \includegraphics[width=0.45\columnwidth]{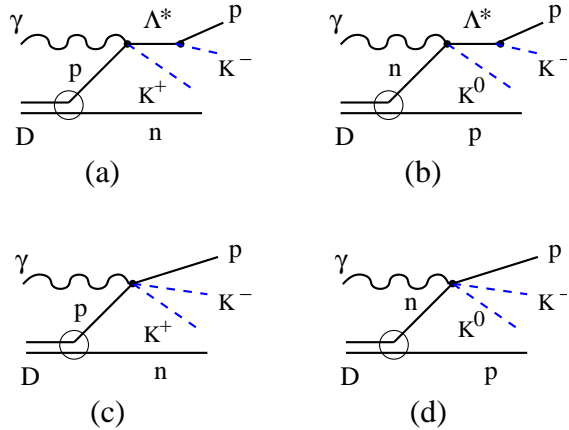}
  \caption{\label{FIG:15}{\small%\tcaps%
  Diagrammatic representation of background spectator channels.
  (a,b): Quasi-free $\Ls K$ photoproduction, (c,d): quasi-free non-resonant
  $pK^-K$ photoproduction.}}}
 \end{figure}

The spectator channels are depicted in Fig.~\ref{FIG:15}.
Contributions of these channels to the invariant mass distribution
are defined by Eqs.~(\ref{E1}) and (\ref{E2}), where the
amplitudes are expressed via products of the amplitude of the
elementary $\gamma N\to N \bar KK$ reactions $A^{\gamma N}(n)$ and
the deuteron wave function $\phi$ as
\begin{equation}\label{S1}
  T_{fi}(n)=\sqrt{2M_D}\sum\limits_{m}\,
   A^{\gamma N}_{m_2;m\lambda}(n)\phi^{m_D}_{m,m_1}({\bf p}),
\end{equation}
where the deuteron wave function is defined in
Eq.~(\ref{gamma-D4}), and the index $n$ corresponds to the
different elementary sub-processes discussed in Sec.~III.
\begin{figure}[th]
 {%\centering
  \includegraphics[width=0.3\columnwidth]{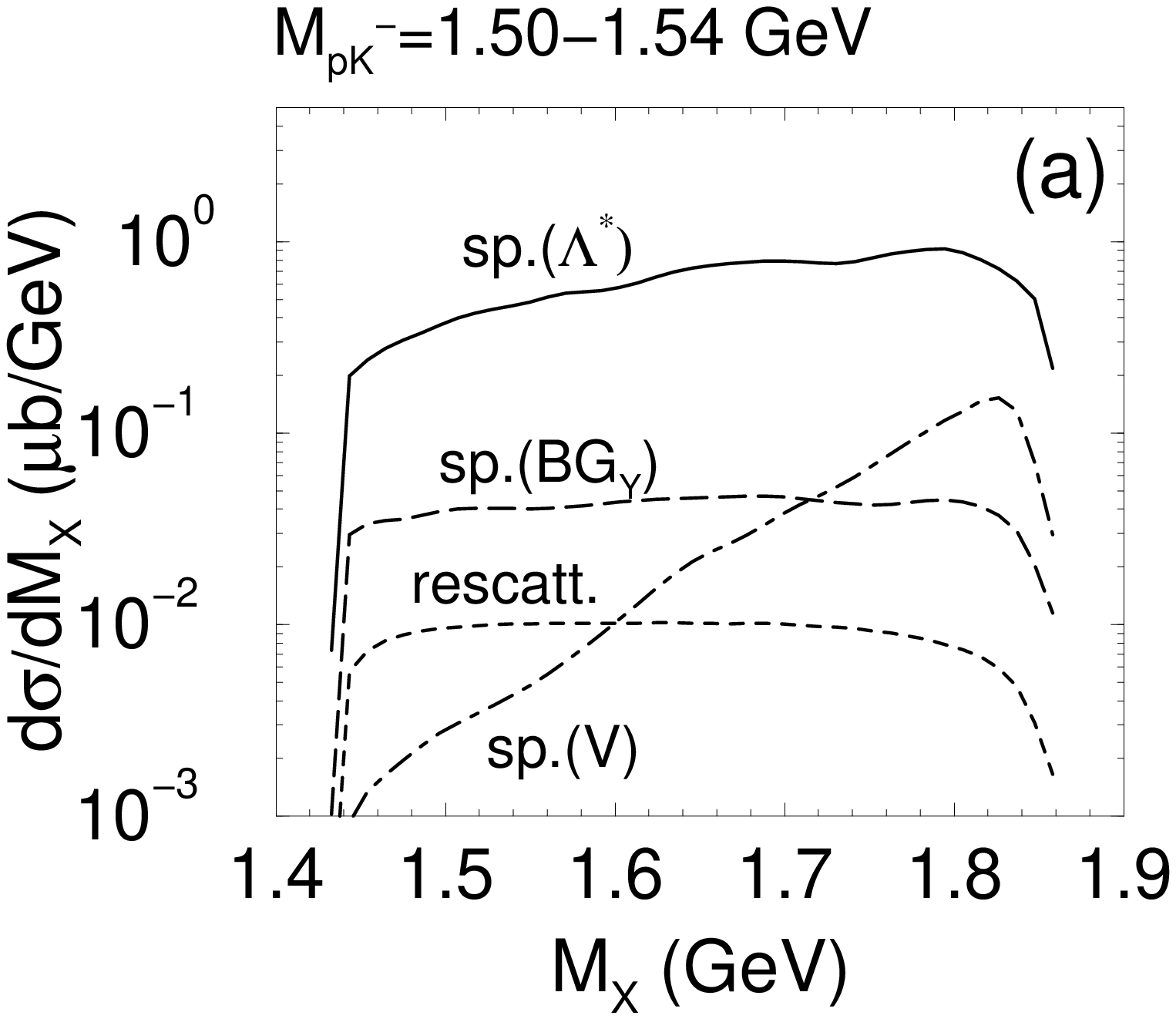}\qquad
  \includegraphics[width=0.3\columnwidth]{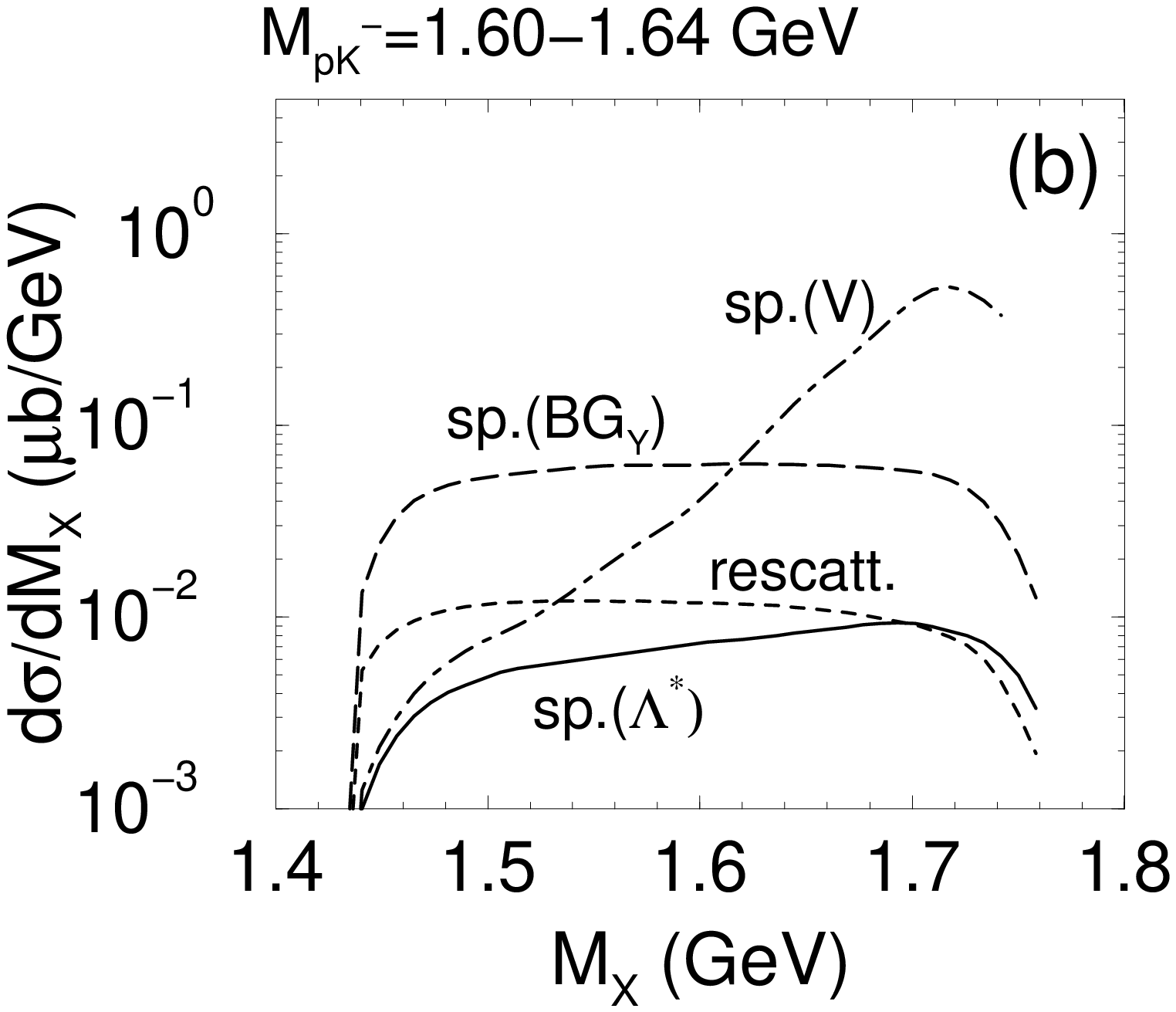}
  \caption{\label{FIG:16}{\small%\tcaps%
Background contributions to the missing-mass distribution in the
$\gamma D\to pK^-X$ reaction at $E_\gamma=2.1$~GeV for quasi-free
$\Ls$ photoproduction, vector meson and hyperon excitations
(spectator mechanism), and rescattering channels shown by solid,
dot-dashed, long-dashed, and dashed curves, respectively. (a) and
(b) correspond to $M_0=1.52$ and 1.62  GeV, respectively.}}}
 \end{figure}
The background contributions for the quasi-free $\Ls$
photoproduction, vector meson and hyperon excitation are shown in
Fig.~\ref{FIG:16} by solid, dot-dashed, and long-dashed curves,
respectively. If the $pK^-$ invariant mass is close to the $\Ls$
mass (cf.\ Fig.~\ref{FIG:16}(a)) the quasi-free $\Ls$
photoproduction gives the dominant contribution to the background.
The next important contribution comes from the $BG_Y$ channel,
parameterized by the constant matrix element. The contribution of
the vector mesons in the region where the $[\gamma D, pK^-]$
missing mass is around the $\Tp$ mass is rather small. Indeed, it
does not contribute to the $\gamma n$ reaction and, moreover, it
is suppressed dynamically. Thus,  the kinematics of the associated
$\Ls\Tp$ photoproduction in the forward direction requires a fast
$K^-$ and slow $K^+$. In this case, the $K^+K^-$ invariant mass is
far from the $\phi$ meson mass. But the situation changes at large
values of the $[\gamma D, pK^-]$ missing mass. In this case the
available values of $K^+K^-$ invariant mass cover the $\phi$ meson
mass region, and the contribution of the $\phi$ meson excitation
becomes essential. The case when the $pK^-$ invariant mass is far
from the $\Ls$ mass is shown in Fig.~\ref{FIG:16}~(b). Now, the
quasi-free $\Ls$ photoproduction is suppressed, whereas the other
channels have the same order of magnitude.

\subsection{Rescattering processes}

\begin{figure}[th!]
{\centering
 \includegraphics[width=.5\textwidth]{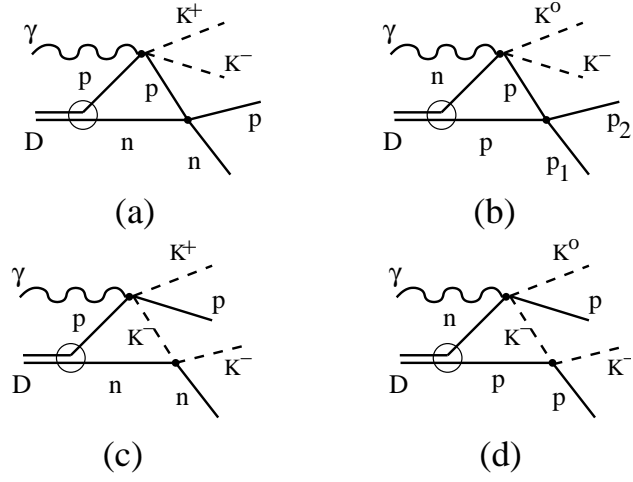}
 \caption{\label{FIG:17}{\small%\tcaps%
 Rescattering of the proton (a,b) and $K^-$ meson (c,d)
 in the $pK^-NK$ photoproduction from the deuteron.}}}
\end{figure}

 Let us consider  proton and  $K^-$ meson rescattering when
 a $pK^-$ pair is produced in the $\gamma N$ interaction as shown in
 Fig.~\ref{FIG:17}~(a,b) and (c,d), respectively.
 In principle, the
 $K^+n$ and $K^0p$  rescattering in quasi-free $\gamma N\to pK^-K$
 photoproduction from a deuteron must be taken into account too, but we
 skip them at the present stage, assuming first that such processes give
 a small correction to the spectator (quasi-free photoproduction)
 channels considered in the previous section and second, the cross
 section for elastic $K^+N$ scattering is much smaller than
 for $K^-N$ scattering~\cite{PDG}. Moreover, as we will show,
 the dominant contribution here comes from proton
 rescattering and, therefore,  kaon rescattering is a small part.
Basically, the amplitudes of the rescattering processes are
evaluated similarly to the amplitudes of the associated coherent
$\Tp\Ls$ or $\Tp pK^-$ photoproduction considered  in Sect.~III,
where we assumed the dominance of the imaginary part of the
corresponding loop diagrams, calculated by cutting rules. But
there are several
new aspects:\\
 First, the rescattered particles are outside of
 the production plane and, therefore now, in the loop integrals we
have to integrate over the virtual nucleon momentum $p$ and the
azimuthal angle $\varphi_p$. If the polar and azimuthal angles of
the momentum ${\bf p}_\xi$ are $\theta_\xi$ and  $\varphi_\xi$,
respectively,  then the three dimensional vector of the virtual
nucleon in the laboratory frame, with the $z$-axis along the beam
direction, reads
\begin{eqnarray}
 p_x&=&p_x'\cos\varphi_\xi - p_y'\sin\varphi_\xi~,\nonumber\\
 p_y&=&p_y'\cos\varphi_\xi + p_x'\sin\varphi_\xi~,\nonumber\\
 p_z&=&p(\cos\theta_p\cos\theta_\xi
 -\sin\theta_p\sin\theta_\xi\cos\varphi_p)~,\label{R1}
\end{eqnarray}
where
\begin{eqnarray}
 p_x'&=&p(\sin\theta_p\cos\varphi_p\cos\theta_\xi +
 \cos\theta_p\sin\theta_\xi)~,\nonumber\\
 p_y'&=&p\sin\theta_p\sin_p\varphi_p~.\nonumber
\end{eqnarray}
The polar angle $\theta_p$ is fixed by the on-shell conditions
(cf. Eq.~(\ref{cosine}))
\begin{equation}\label{R2}
 a(p,{\bf p}_\xi)\equiv \cos\theta_p =
 \frac{M_K^2-M_\xi^2-M_N^2 +2E_\xi E_p}{2|{\bf p}||{\bf p}_\xi|}~,
\end{equation}
where the four momenta $p_\xi=p_{(\alpha,\beta)}$ are defined as follows.\\
$p$ rescattering:
\begin{eqnarray}\label{R3}
p_\alpha&=&p_{N}+p_f,\qquad p_\beta=p_{K} + p_{K^-} - k_\gamma~,
\end{eqnarray}
$K^-$ rescattering:
\begin{eqnarray}\label{R4}
p_\alpha&=&p_{N}+p_{K^-},\qquad
p_\beta=p_f + p_{K}-k_\gamma~,\nonumber\\
\end{eqnarray}
were $p_N$ and $p_f$ refer to the momenta of the outgoing nucleon
and the proton of the $X$ and $Y$ systems, respectively.
 We remind the reader that the indices $\alpha,\beta$ refer to the cut
loops shown in Fig.~\ref{FIG:9}~($\alpha$) and ($\beta$),
respectively.
\begin{figure}[th]
 {%\centering
  \includegraphics[width=0.3\columnwidth]{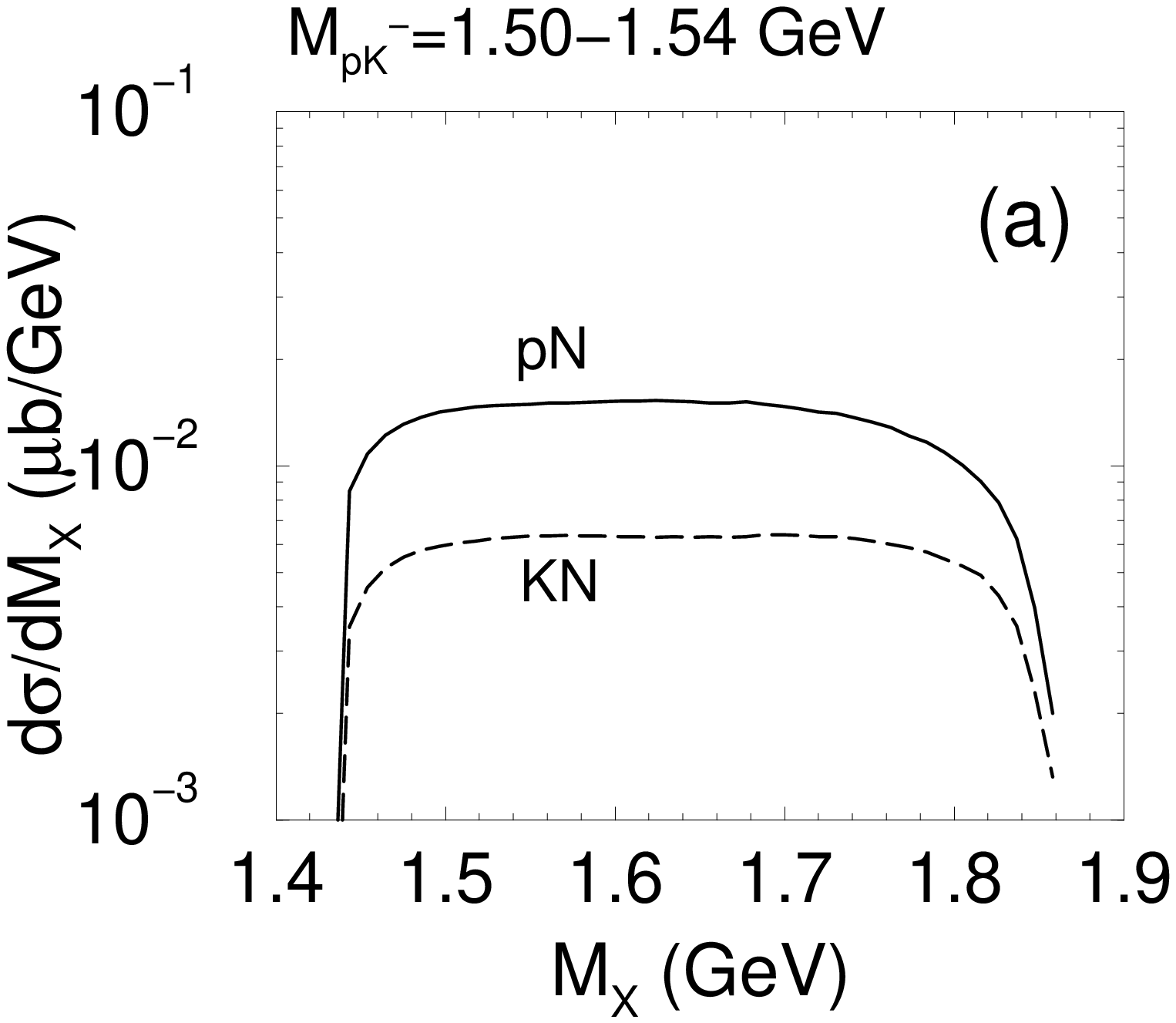}\qquad
  \includegraphics[width=0.3\columnwidth]{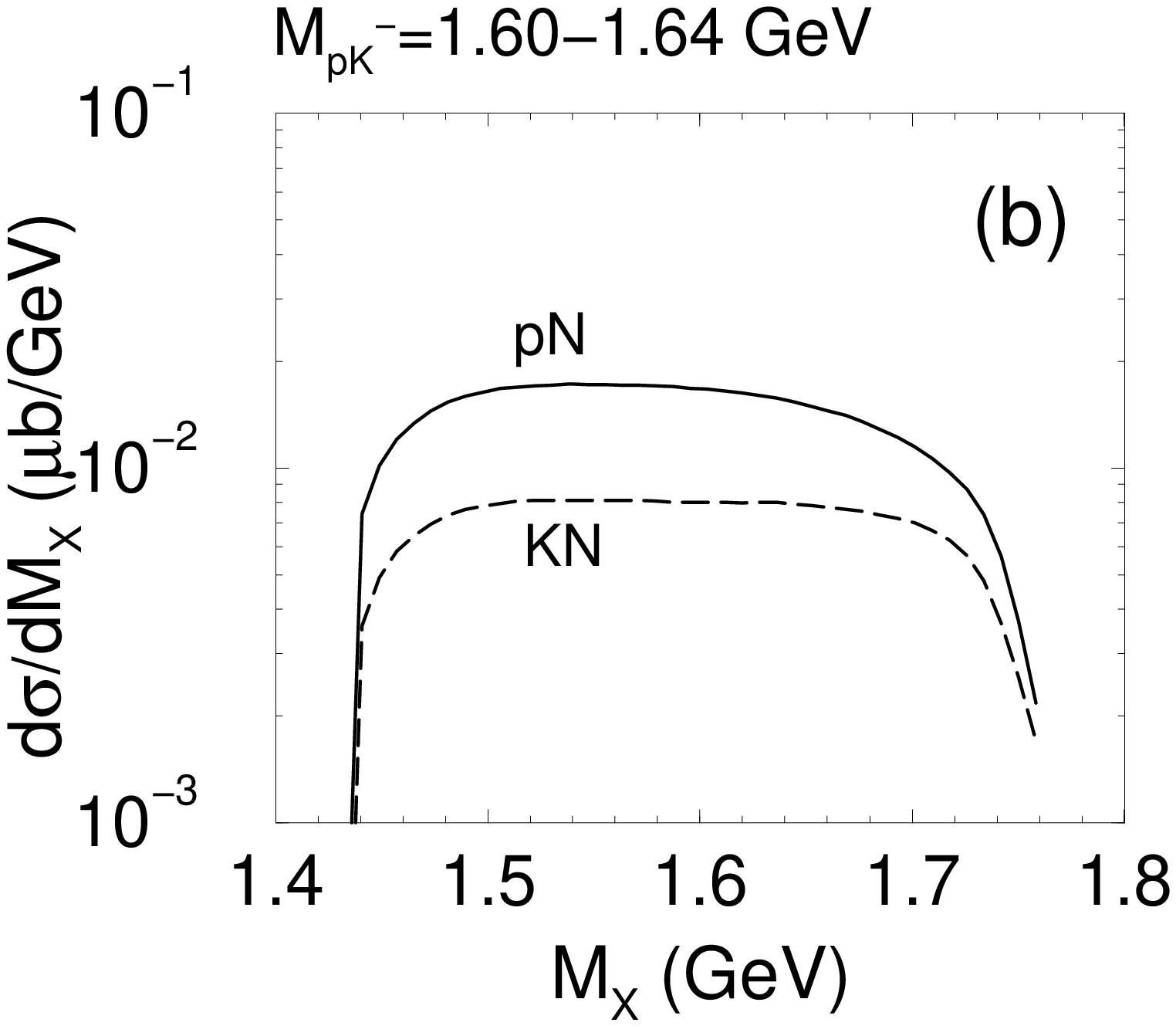}
  \caption{\label{FIG:18}{\small%\tcaps%
The $pN$ and $K^-N$ rescattering channels in $\gamma D\to pK^-X$
reaction at $E_\gamma=2.1$~GeV  for $M_0=1.52$ and 1.62 GeV shown
at (a) and (b), respectively. }}}
 \end{figure}
 To preserve the energy-momentum conservation in the loop vertices
 one has to be careful with the determination of the azimuthal angles
 $\varphi_\xi$.
 The corresponding expression reads
 \begin{eqnarray}
 &&\varphi_\xi=\bar\varphi_\xi\theta(s_\xi) +
 (2\pi - \bar\varphi_\xi)\theta(-s_\xi)~,\nonumber\\
 &&\cos\bar\varphi_\xi=\frac{{p_\xi}_x}{|{\bf p}_\xi|\sin\theta_\xi}~,
 \qquad
 s_\xi=\frac{{p_\xi}_y}{|{\bf p}_\xi|\sin\theta_\xi}~.
 \end{eqnarray}

 Next, one has to choose the effective amplitudes in the loop
integrals. We take them as a product of the deuteron wave
function, photoproduction of the $K^- K$ pair in $\gamma N\to
pK^-K$ reaction and elastic scattering amplitudes. For the proton
and $K^-$meson rescattering they read correspondingly
 \begin{eqnarray}
 \label{R5}
 T_{m_pm_N;m_D\lambda}^{p}&=&\sqrt{2M_D}
 \sum\limits_{mm_1m_2}
 T_{m,m_1\lambda}^{\gamma N\to pKK^-}
 \, T_{m_pm_N,m m_2}^{pN\to pN}
 \,\phi^{m_D}_{m_1m_2}(p,a(p,{\bf p}_\xi))~,\nonumber\\
 T_{m_pm_N;m_D\lambda}^{K^-}&=&\sqrt{2M_D}
 \sum\limits_{m_1m_2}
 T_{m_p,m_1\lambda}^{\gamma N\to pKK^-}
 \, T_{m_N,m_2}^{K^-N\to K^-N}
 \,\phi^{m_D}_{m_1m_2}(p,a(p,{\bf p}_\xi))~,
 \end{eqnarray}
 where $m_p$ and $m_N$ are the spin projections of the outgoing
 proton and nucleon, respectively.
 In our calculations we use an on-shell approximation for the elastic
 scattering amplitudes $T_{fi}^{pN\to pN}$ and
 $ T_{fi}^{K^-N\to K^-N}$, taken from the experimentally
 measured cross sections of elastic scattering. Details of
 the employed parameterizations are given in Appendix~B.

 An important point is related to the $\gamma N\to pKK^-$
 vertex. In our model it consists of three components:
 $\Ls$ excitation, the vector meson contribution, and
 the remaining background contribution denoted above
 as $BG_Y$. Consider first the $\Ls$ channel.  Due to the rescattering
 kinematics, the invariant mass of the virtual $pK^-$ pair
 in the loop integrals covers the $\Ls$ resonance region
 even when the invariant mass of the outgoing $p$
 and $K^-$ meson is outside the resonance position.
 This results in increasing the background contribution
 at $M_0\neq M_\Ls$. But the situation is not so simple.
 Since the $\Ls$ has a small total decay width, $\Gamma_{\rm tot}\simeq
 15.6$~MeV, its decay length is large,
\begin{equation}\label{R6}
  l_0 = {v}{t_0} = \frac{v}{c}\frac{\hbar c}{\Gamma_{\rm tot}}\simeq
  6-10~{\rm fm},
\end{equation}
at a $\Ls$ velocity ${v}\sim (0.5-0.8)\,c$. In other words, the
$\Ls$ decays mostly outside of the deuteron. Similar or even
larger suppression is expected for the case of the $\phi$-meson
contribution. Therefore, we can simply neglect these two channels.

In Fig.~\ref{FIG:18} we show the relative contributions of $p$ and
$K^-$ rescattering for different channels. Enhancement of the
$p$-rescattering is explained by the difference between the cross
sections for $pN$ and $K^-N$ elastic scattering.

\begin{figure}[th]
 {%\centering
  \includegraphics[width=0.3\columnwidth]{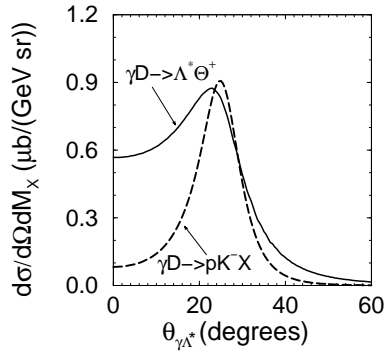}\qquad
  \caption{\label{FIG:19}{\small%\tcaps%
The angular dependence of the missing mass distribution for the
associated $\Ls\Tp$ (solid curve) photoproduction and the
background spectator processes (dashed curve) at $M_X=1.53$~GeV,
$M_0=1.52$~GeV, and $E_\gamma=2.1$~GeV. }}}
 \end{figure}

To summarize this section we conclude that at low energy, say
$E_\gamma=1.7-2.3$~GeV, the dominant component of the non-resonant
background comes from the quasi-free $\Ls$ spectator channel, the
next important contribution is composed of the $BG_Y$ spectator
channel and $pN$ rescattering, if the invariant mass of the $pK^-$
pair is inside  the $\Ls$ resonance position with $M_0=1.52$~GeV.
Outside of the $\Ls$ resonance position the quasi-free  $\Ls$
spectator channel is strongly suppressed, whereas other channels
remain on the same level

Finally, let us examine the angular dependence of the spectator
channel, similar to  the associated $\Ls\Tp$ photoproduction (cf.
Figs.~\ref{FIG:13} and \ref{FIG:14}). The corresponding result is
shown in Fig.~\ref{FIG:19} where we present simultaneously
contributions from the associated $\Ls\Tp$ photoproduction
(signal) and from the background (noise) dominated by the
spectator channels. The calculation is for the resonance region
with $M_X=1.53$~GeV, $M_0=1.52$~GeV, and $E_\gamma=2.1$~GeV. We
choose the case of $J_\Theta^P=3/2^-$. One can see that the
spectator channel has a sharp peak caused by the deuteron wave
function with a maximum close to the maximum for coherent $\Ls\Tp$
photoproduction. At small angles the noise decreases much faster
than the signal. Therefore, we can conclude that the largest value
for the S/N ratio is expected at extremely forward $pK^-$
photoproduction angles, say $\theta_{\rm c.m.s.}\le 22^0$.

\section{Results and discussion}

Below, we discuss the prediction for the $\Tp$ formation processes
under two different conditions. The first is  $\Tp$
photoproduction at low energy ($E_\gamma=1.7-2.3$~GeV) in the
inclusive $\gamma D\to pK^-X$ reaction. It can be studied, for
example, by LEPS at SPring-8 and/or Crystal-Barrel at ELSA (Bonn).
The second is the formation $\Tp$ photoproduction process in
exclusive $\gamma D\to pK^-nK^+$ reaction in a wider energy
interval ($E_\gamma=1.7-3.5$~GeV) with the experimental conditions
of the CLAS Collaboration measurement~\cite{JLab-06}. All
calculations are made for a total $\Tp$ decay width of
$\Gamma_\Theta=1$~MeV.

\subsection{The missing mass distribution in inclusive
$\gamma D\to pK^-X$ reactions}

We calculate the missing mass distribution in the inclusive
$\gamma D\to pK^-X$ reaction with two cuts. The first one is the
$\phi$-meson cut. We exclude all events with a $K^+K^-$ invariant
mass close to the mass of the $\phi$ meson: $\left|
M_{K^+K^-}-M_\phi\right| < 20$~MeV~\cite{Nakano03}. The second one
is the angular cut: we keep only forward $pK^-$ pair
photoproduction with $\theta_{\rm c.m.s.}\le22^0$. This cut gives
a maximum  S/N ratio. Hereafter, we define the corresponding
missing mass distribution as ${d\sigma^F}/{dM_X}$, where the
superscript $F$ indicates, conditionally, an alignment in the
forward photoproduction of the $pK^-$ pair.
\begin{figure}[th]
 {%\centering
  \includegraphics[width=0.3\columnwidth]{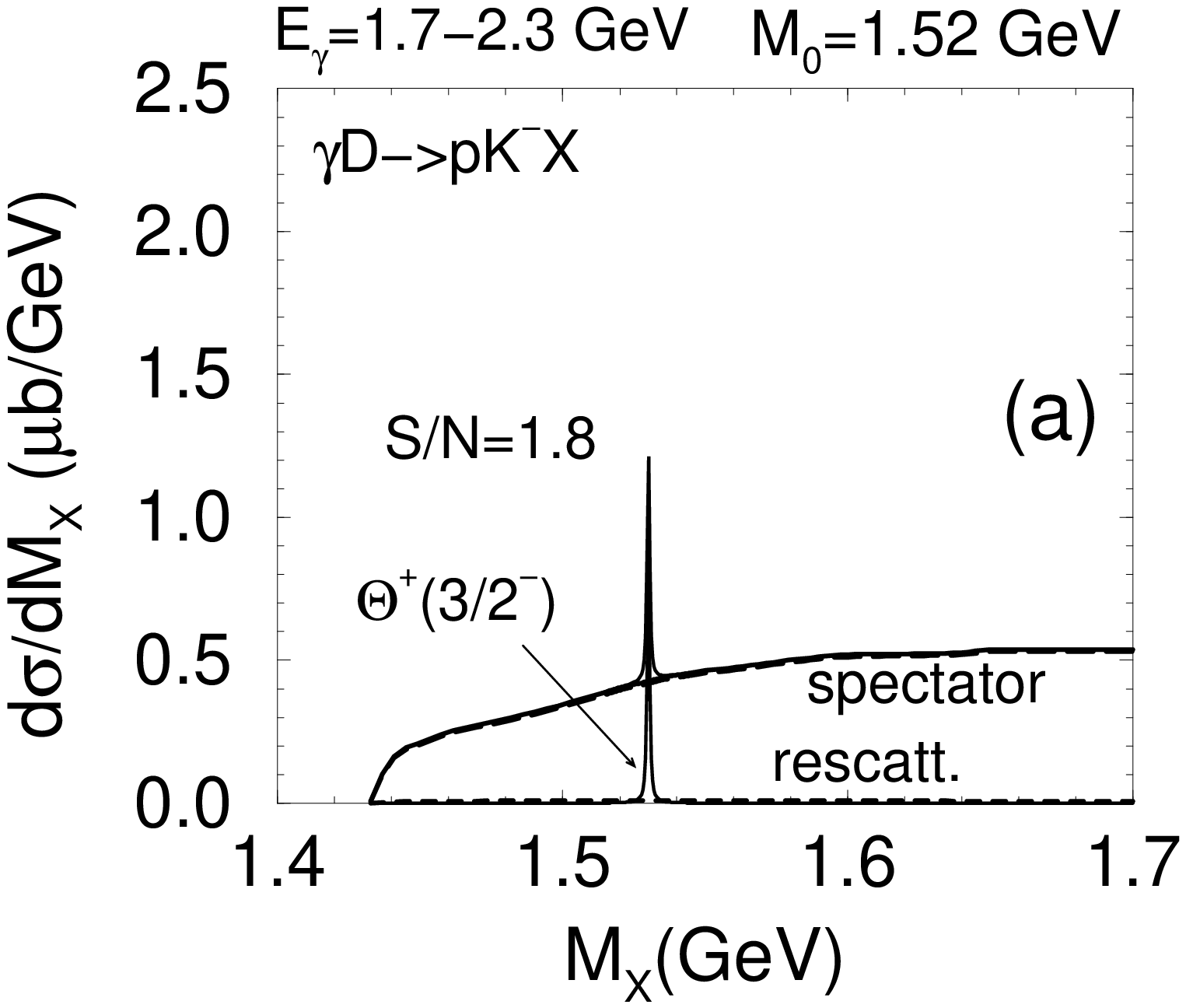}\qquad
  \includegraphics[width=0.3\columnwidth]{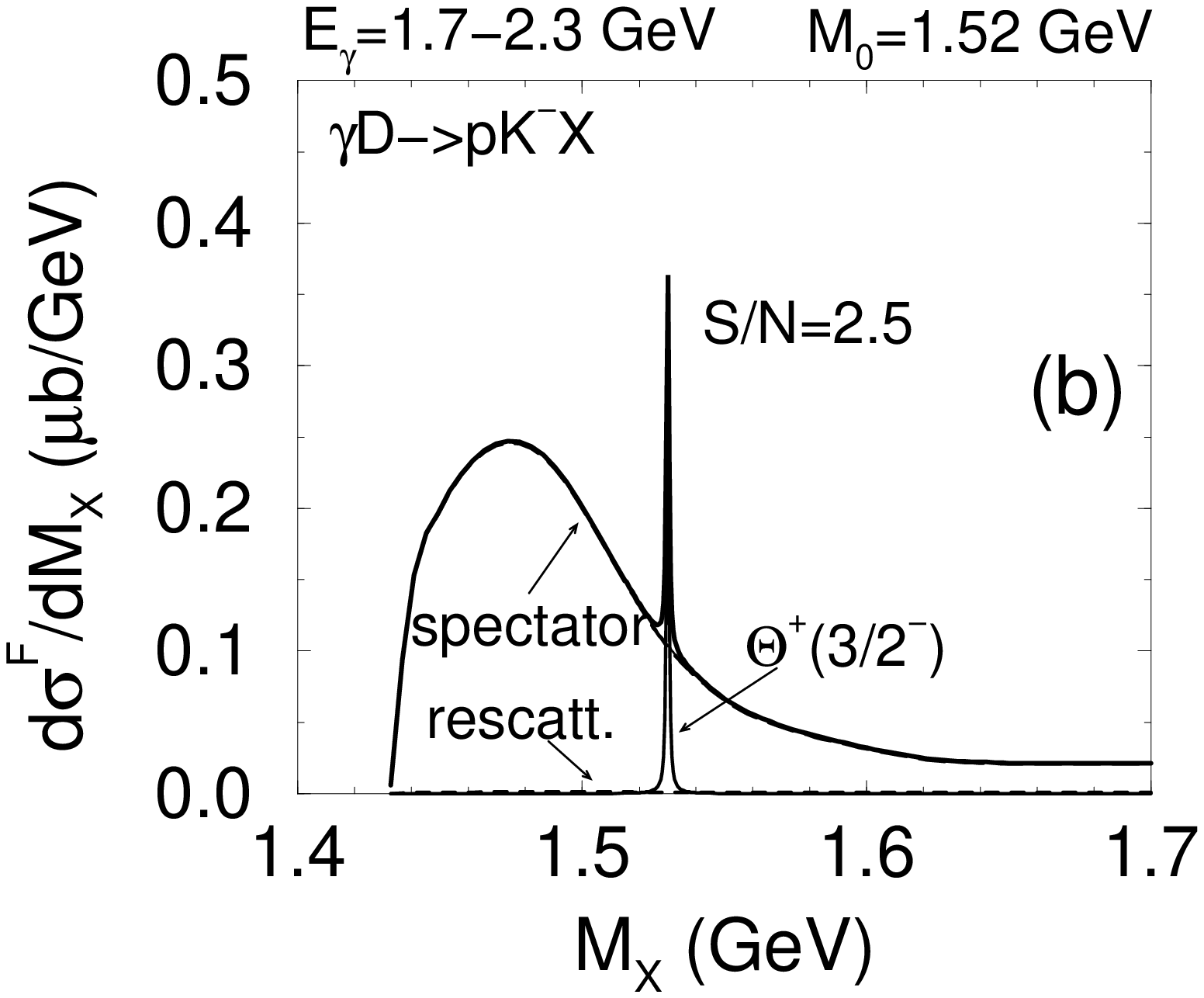}
  \caption{\label{FIG:20}{\small%\tcaps%
 The $[\gamma D,pK^-]$ missing mass distribution in the $\gamma D\to  pK^-X$
 reaction together with the partial contributions of different
 background channels at $M_0=1.52$~GeV and $E_\gamma=1.7-2.3$~GeV.
 The contributions from the $\Tp$ signal, spectator and rescattering processes
 are shown by the thin, long dashed, and dashed curves, respectively.
(a) and (b)
 correspond to calculations without and with the angular cut $\theta_{pK^-}\le22^0$,
 respectively.}}}
 \end{figure}
 Figure~\ref{FIG:20} illustrates the effect of the angular cut
 in the missing mass distribution $\gamma D\to  pK^-X$
 reaction at the $\Ls$ resonance position with $M_{pK^-}\sim M_\Ls$
 ($M_0=1.52$~GeV), averaging over the energy interval
 $E_\gamma=1.7-2.3$~GeV. Here, we choose the case of $J_\Theta^P=3/2^-$.
  First of all, in the $\Ls$ resonance region one can see
  a distinct effect of the
 associated $\Ls\Tp$ photoproduction as a sharp $\Tp$ peak against
 the flat non-resonant background with and without the angular cut.
 The angular cut increases the S/N ratio.
 Fig.~\ref{FIG:20}~(a)
 shows the missing mass distribution calculated
 for $\theta_{pK^-}\le\pi/2$ (c.m.s.),
 whereas in Fig.~\ref{FIG:20}~(b) the angular interval is limited to
 $22^0$ in accordance with the results in Fig.~\ref{FIG:19}. The angular
 cut suppresses the  cross sections of both the $\Tp$ signal
 and the background; it further modifies the shape
 of the background in such a way
 as to get the maximum S/N ratio at the resonance position $M_X=M_\Theta$.
 The shape modification is explained by the suppression of the quasi-free
 spectator channel at large invariant mass $M_X$. When $M_X$ increases
 the maximum in the quasi-free background distribution, shown in
 Fig.~\ref{FIG:19} (dashed curve), moves towards large angles
 being outside of the integration region and, therefore, the
 contribution of this channel decreases.
 As a result, the angular cut creates
 an enhancement of the S/N ratio by about $40\%$.

\begin{figure}[th]
 {%\centering
  \includegraphics[width=0.3\columnwidth]{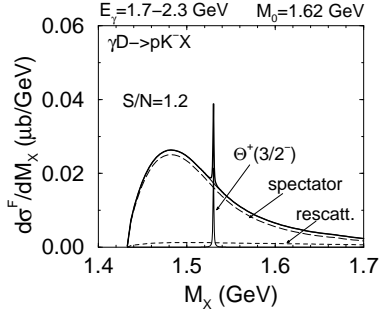}
  \caption{\label{FIG:21}{\small%\tcaps%
 The same as in Fig.~\protect\ref{FIG:20}~(b), but for $M_0=1.62$~GeV.}}}
 \end{figure}

Figure~\ref{FIG:21} illustrates the case when the  $pK^-$
invariant mass is far from the $\Ls$ resonance position,
$M_0=1.62$~GeV. Now we also see some $\Tp$ peak generated mainly
by the associated non-resonant processes. The resonance channel is
suppressed because of the small $\Ls$ decay width. The background
near the $\Tp$ peak is dominated by the $BG_Y$ and rescattering
channels. Since now the charged and neutral $K$ meson exchange
diagrams contribute incoherently, the ratio $\rm (S/N)_{NR}$ would
be smaller compared to the ratio in the resonance region, $\rm
(S/N)_{R}$. Neglecting the rescattering channels and the
background shape modification, one can get the following
qualitative estimate
\begin{equation}\label{RES1}
  \rm \left(\frac{S}{N}\right)_{NR}
  \simeq \rm \frac12\,\left(\frac{S}{N}\right)_{R}~.
\end{equation}
The background shape modification and the rescattering channels
result in decreasing $\rm (S/N)_{NR}$.

\begin{figure}[th]
 {%\centering
 \includegraphics[width=0.21\columnwidth]{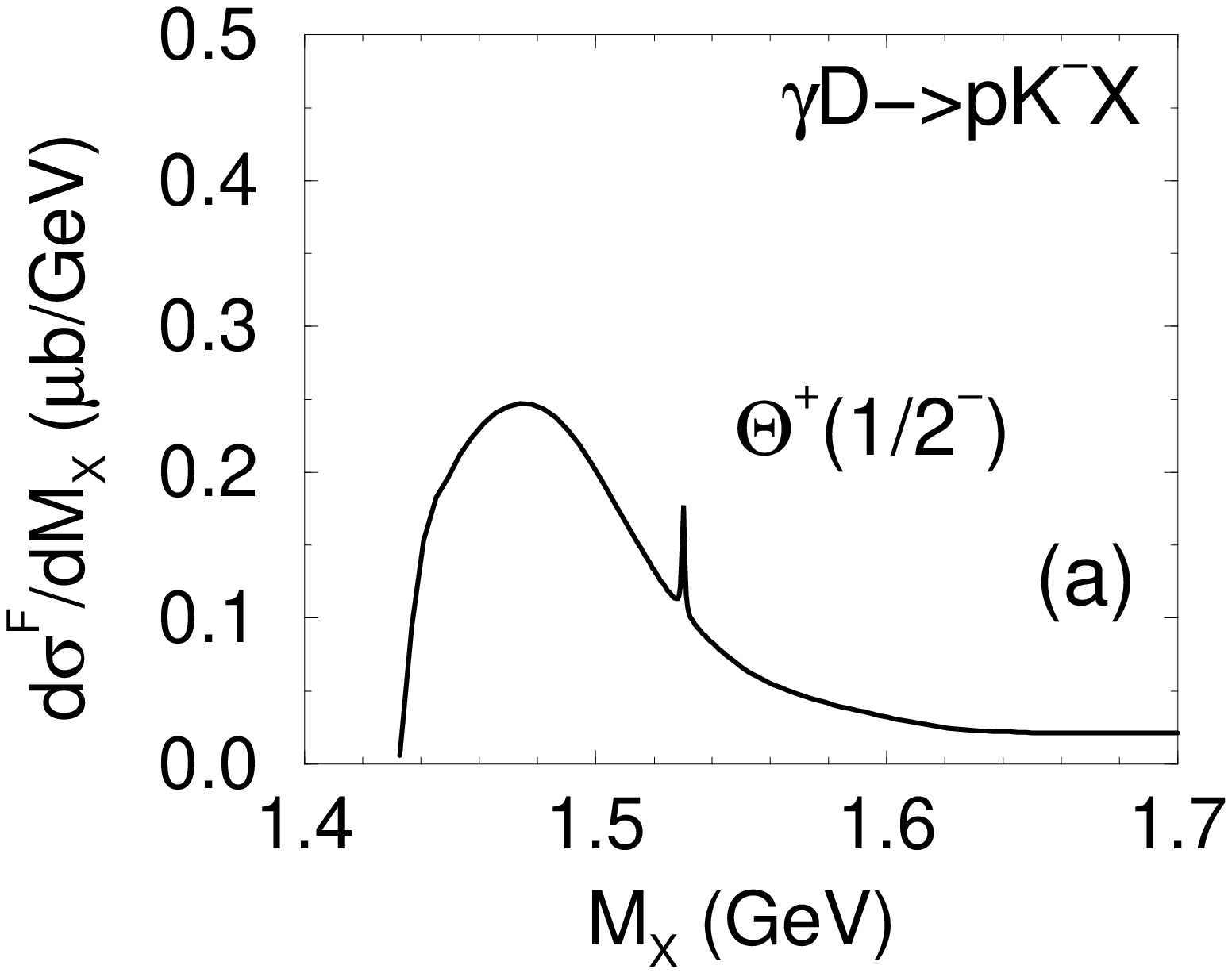}\qquad
  \includegraphics[width=0.21\columnwidth]{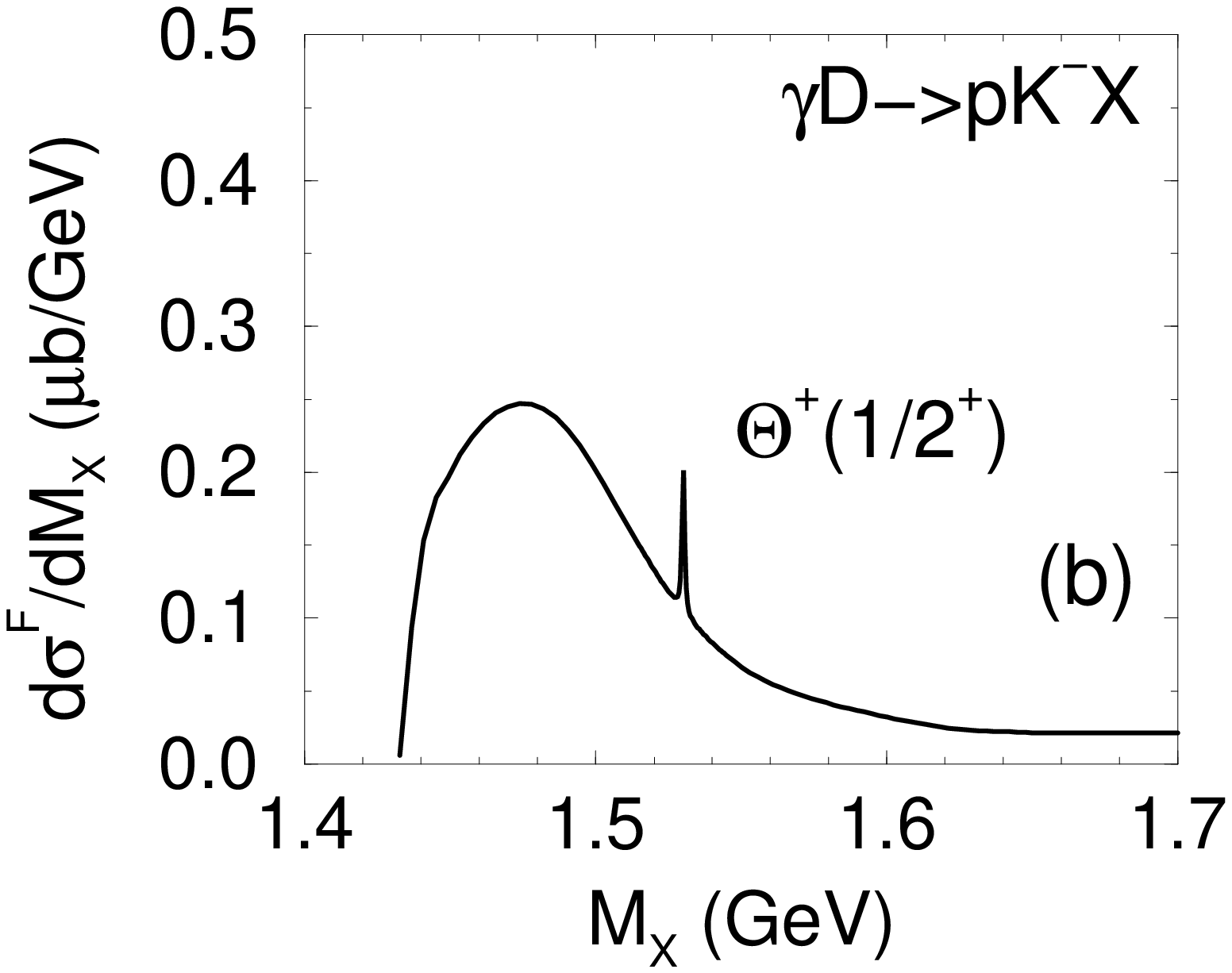}\qquad
  \includegraphics[width=0.21\columnwidth]{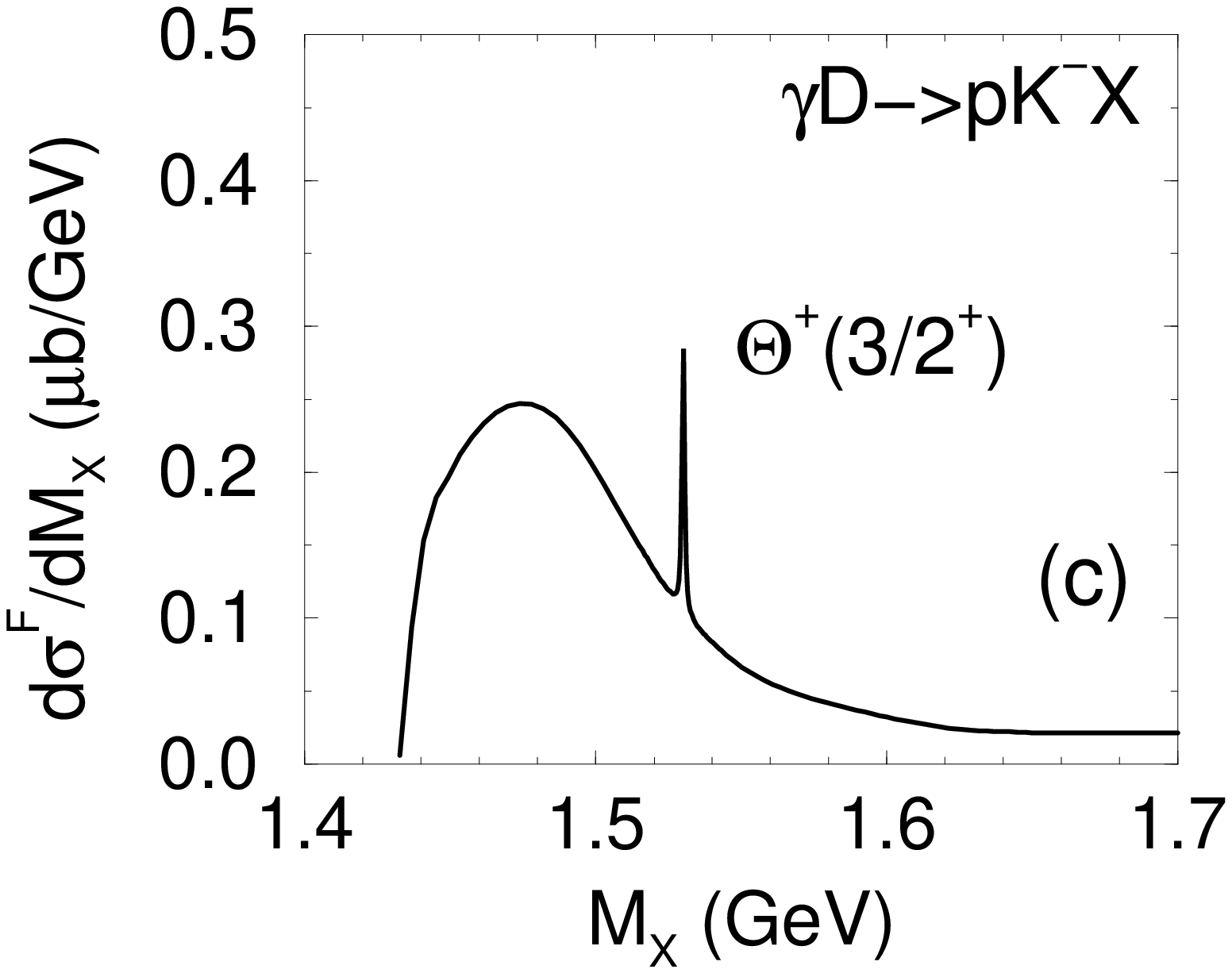}\qquad
  \includegraphics[width=0.21\columnwidth]{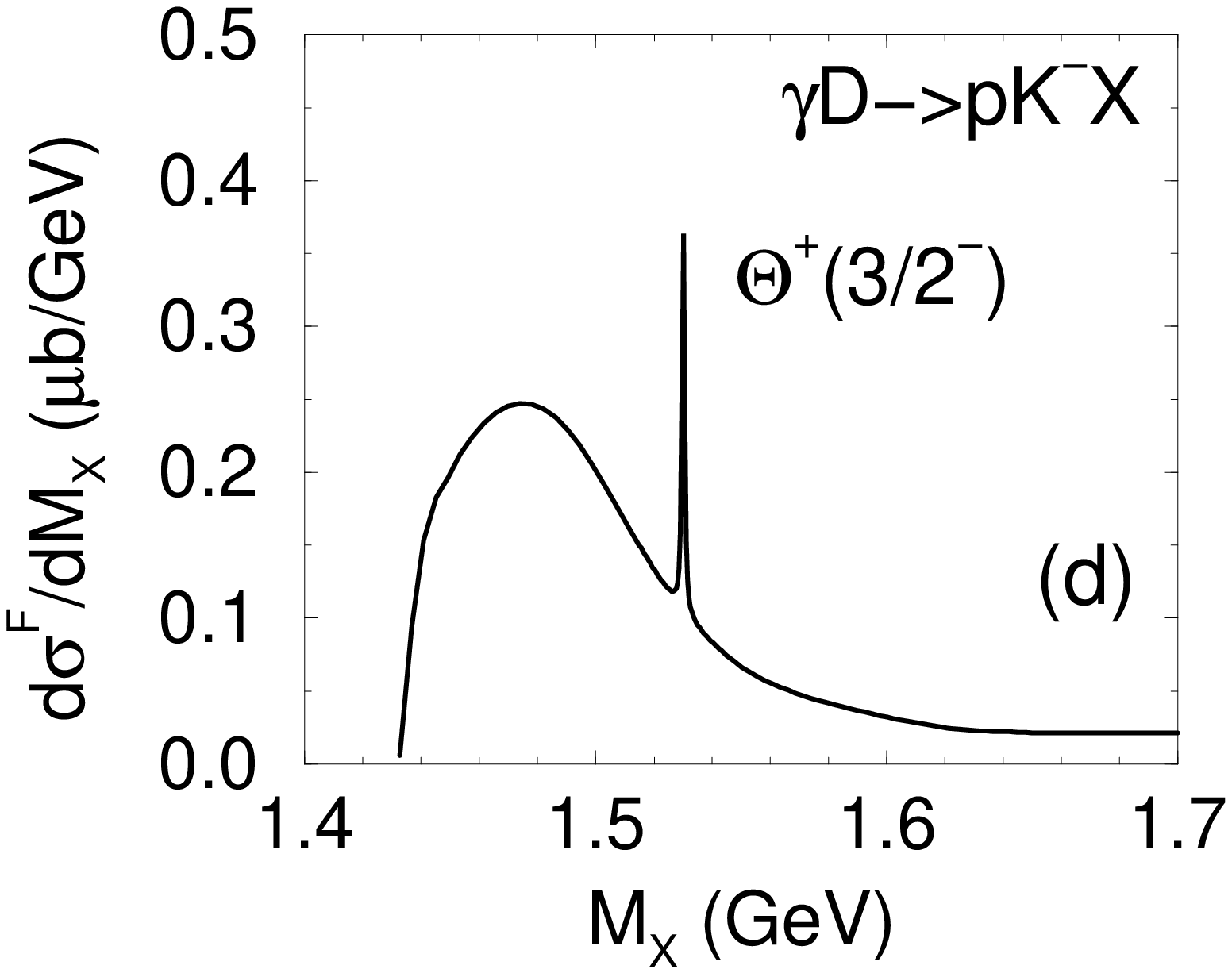}
  \caption{\label{FIG:22}{\small%\tcaps%
 Missing mass distribution in the $\gamma D\to  pK^-X$
 reaction  at $E_\gamma=1.7-2.3$~GeV and $M_0=1.52$~GeV for
 different $\Tp$ spin and parity: (a), (b), (c) and (d)
 correspond to the $\Tp$ spin and parity  $J^P=\frac12^-,\,
\frac12^+,\,\frac32^+$, and $\frac32^-$, respectively. }}}
 \end{figure}

  In Fig.~\ref{FIG:22} we show the
  missing mass distribution for different $\Tp$ spin and parity:
  $J^P=\frac12^\mp,\frac32^\pm$.
  The signal-to-noise ratio for different $J^P$
  reads
\begin{equation}\label{RES111}
  \frac12^-:\frac12^+:\frac32^+:\frac32^-\simeq
  0.7:0.9:1.7:2.5~.
\end{equation}
This result is in qualitative agreement with the previous analysis
of the differential invariant mass distributions  (cf.
Eq.~(\ref{RES11})).

Note that the value of the invariant mass distribution at the
resonance is independent of the $\Tp$ decay width. But since all
experiments have a finite experimental resolution, the measured
signal is smeared by the experimental resolution, and this smeared
signal must depend on the value of $\Gamma_\Theta$.
\begin{figure}[th]
 {%\centering
  \includegraphics[width=0.35\columnwidth]{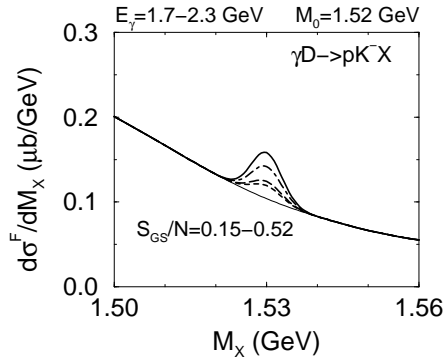}
  \caption{\label{FIG:23}{\small%\tcaps%
  The  missing mass distribution for $\gamma D\to  pK^-NK$
  at $E_\gamma=1.7-2.3$~GeV and $M_0=1.52$~GeV and for different
  $\Tp$ spin and parity, folded with a Gaussian resolution function.}}}
 \end{figure}
  In Fig.~\ref{FIG:23} we show  the missing
  mass distribution folded with a Gaussian distribution function
\begin{eqnarray}\label{RES12}
 \frac{d\sigma}{dM_X}&=&\int
 \frac{d\sigma}{dM}\, f(M_X-M)dM~,\nonumber\\
  f(M_X-M)&=&\frac{1}{\sigma\sqrt{2\pi}}
  \exp\left[-{\frac{(M_X-M)^2}{2\sigma^2}}\right]~,
\end{eqnarray}
with $\sigma=3$~MeV, which imitates a finite experimental
resolution. In this case, the height of the resonance peak (S)
decreases proportional to the factor of
\begin{equation}\label{RES13}
\frac{\sqrt{\pi}}{2\sqrt{2}}\, \frac{\Gamma_\Theta}{\sigma}\simeq
0.21.
\end{equation}
Therefore, in order to see this peak above the background, one
needs rather good experimental resolution even assuming that the
$\Tp$ spin parity is $3/2^-$.

Finally, let us estimate the total $\Tp$ formation cross section
in inclusive $\gamma D\to pK^-X$ reactions with angle and
$\phi$-meson cuts. For $J^P_\Theta=3/2^-$, the $\Tp$
photoproduction cross section at maximum is equal to
\begin{equation}\label{Theta-signalSP}
  \frac{d\sigma^{\Tp\,F}}{dM}{\Big|_{\rm max}}\simeq 0.26\,
  \frac{\mu{\rm b}}{\rm GeV}~,
\end{equation}
and this value is independent of the $\Tp$ decay width. Then, the
total cross section for the $\Tp$ signal reads
\begin{eqnarray}
 \sigma^{\Tp\,F}_{\rm tot}
 \simeq
 \frac{\pi}{2}\times\Gamma_\Theta
 \times
 \frac{d\sigma^{\Tp\,F}}{dM}{\Big|_{\rm max}}
 \simeq  0.41~{\rm nb}~.
\end{eqnarray}
 Results for other assignments  of $J_\Theta^P$ can be evaluated using
 Eq.~(\ref{RES111}).

\subsection{$\bm{\Theta^+}$
formation processes in exclusive ${\bm{\gamma d\to pK^-nK^+}}$
reaction}

The reaction $\gamma D\to \Tp pK^-\to npK^+K^-$ has been analyzed
recently by the CLAS collaboration~\cite{JLab-06}. We note that
this experiment was designed for studying the $\Tp$
photoproduction in direct $\gamma n\to \Tp K^-$ elementary
processes. In order to enforce the $\Tp$ signal the data analysis
was performed with some specific cuts. For convenience, we denote
those cuts together with acceptance of the CLAS detector as the
CLAS experimental conditions. No sizable $\Tp$ signal in the
$nK^+$ invariant mass distribution was observed. Therefore, it
seems to be interesting and important to estimate the cross
section for $\Tp$ formation  for the condition of this experiment.
If we find that the formation cross section is greater than the
experimental accuracy of the CLAS experiment, then it indicates
problems for the $\Tp$: either $\Tp$ does not exist, or it does
exist but the $\Tp$ width is much smaller than 1~MeV.
\begin{figure}[th]
 {%\centering
  \includegraphics[width=0.35\columnwidth]{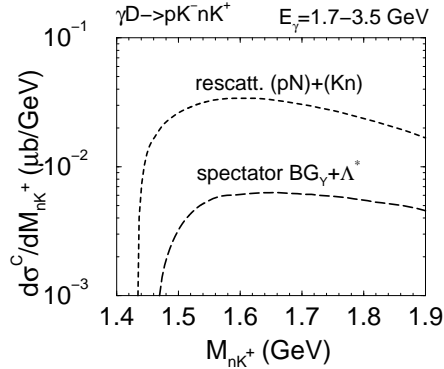}
  \caption{\label{FIG:24}{\small%\tcaps%
   The non-resonant background structure of the $nK^+$ invariant mass
   distribution in $\gamma D\to  pK^-NK$ at $E_\gamma=1.7-3.5$~GeV
   and for CLAS conditions.}}}
 \end{figure}

The acceptance of the CLAS detector allows to detect (i) proton
and kaons with momenta greater than 0.35 and 0.25~(GeV/c),
respectively; (ii) the angles of the direction of flight of the
positively and negatively charged particles are greater than 9$^0$
and 15$^0$ (laboratory system), respectively. The data analysis
was performed with\\
 (iii) $\phi$-meson cut $M_{K^+K^-}> M_c=1.06$~GeV,\\
 (iv) $\Ls$ cut $|M_{pK^-}-M_\Ls|>\Delta_c=25\,{\rm MeV}$,\\
 (v) neutron momentum cut $p_n>p_c=0.2$~GeV/c.

The $\Ls$ cut almost kills the associated $\Ls\Tp$ photoproduction
shown in Fig.~\ref{FIG:1}. "Almost" means that at $\Delta_c \simeq
 25-50$~MeV the $\Ls$ signal is rather weak, but finite.
Nevertheless, the main contribution to the $\Tp$ formation comes
from the non-resonant channels shown in Fig.~\ref{FIG:12}.

The neutron momentum cut strongly reduces the spectator processes,
shown in Fig.~\ref{FIG:15}, making  rescattering channels
dominant. In Fig.~\ref{FIG:24} we show the non-resonant background
for $nK^+$ invariant mass distribution accounting for the CLAS
experimental conditions. One can see the dominance of the
rescattering (mainly $pn$ rescattering) channel. Two other cuts
are dangerous for the formation processes. As we have shown in
Sec.~IV, the dominant contribution to the $\Tp$ formation comes
from the forward photoproduction of a $pK^-$ pair with
$\theta_{pK^-} \lesssim 15^0$ in laboratory system. In this case,
the $K^+$ meson is a slowly moving particle. Therefore, the
acceptance restriction for momentum $p_{K^+}>0.25$~GeV/c together
with the angle limitation for the proton and $K^-$ meson reduce
the cross section of $\Tp$ formation. Note that for the inclusive
photoproduction discussed above the $pK^-$ pairs are detected in
the forward direction, i.e., there are no such restrictions
(cuts), and any value of $p_{K^+}$ consistent with conservation
laws is allowed.

\begin{figure}[th]
 {%\centering
  \includegraphics[width=0.35\columnwidth]{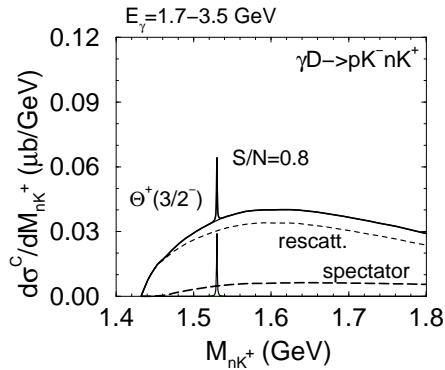}
  \caption{\label{FIG:25}{\small%\tcaps%
   The $nK^+$ invariant mass
   distribution in $\gamma D\to  pK^-NK$ at $E_\gamma=1.7-3.5$~GeV and for
   conditions (i)-(v).}}}
 \end{figure}

   Fig.~\ref{FIG:25} shows the $nK^+$ invariant mass
   distribution in $\gamma D\to  pK^-nK^+$ at $E_\gamma=1.7-3.5$~GeV
   and for CLAS conditions (i) -(v).
   For convenience, we denote this distribution as
   ${d\sigma^C}/{dM_{nK^+}}$, where the superscript "$C$"
   indicates the CLAS conditions.
   One can see some
   $\Tp$ signal against the non-resonant background dominated
   by the rescattering channels. We have chosen the more favorable case
   of $J^P_\Tp=3/2^-$. However, even in this case the S/N ratio
  is about three times smaller compared to the case shown in
  Fig.~\ref{FIG:20}~(b).
\begin{figure}[th]
 {%\centering
  \includegraphics[width=0.4\columnwidth]{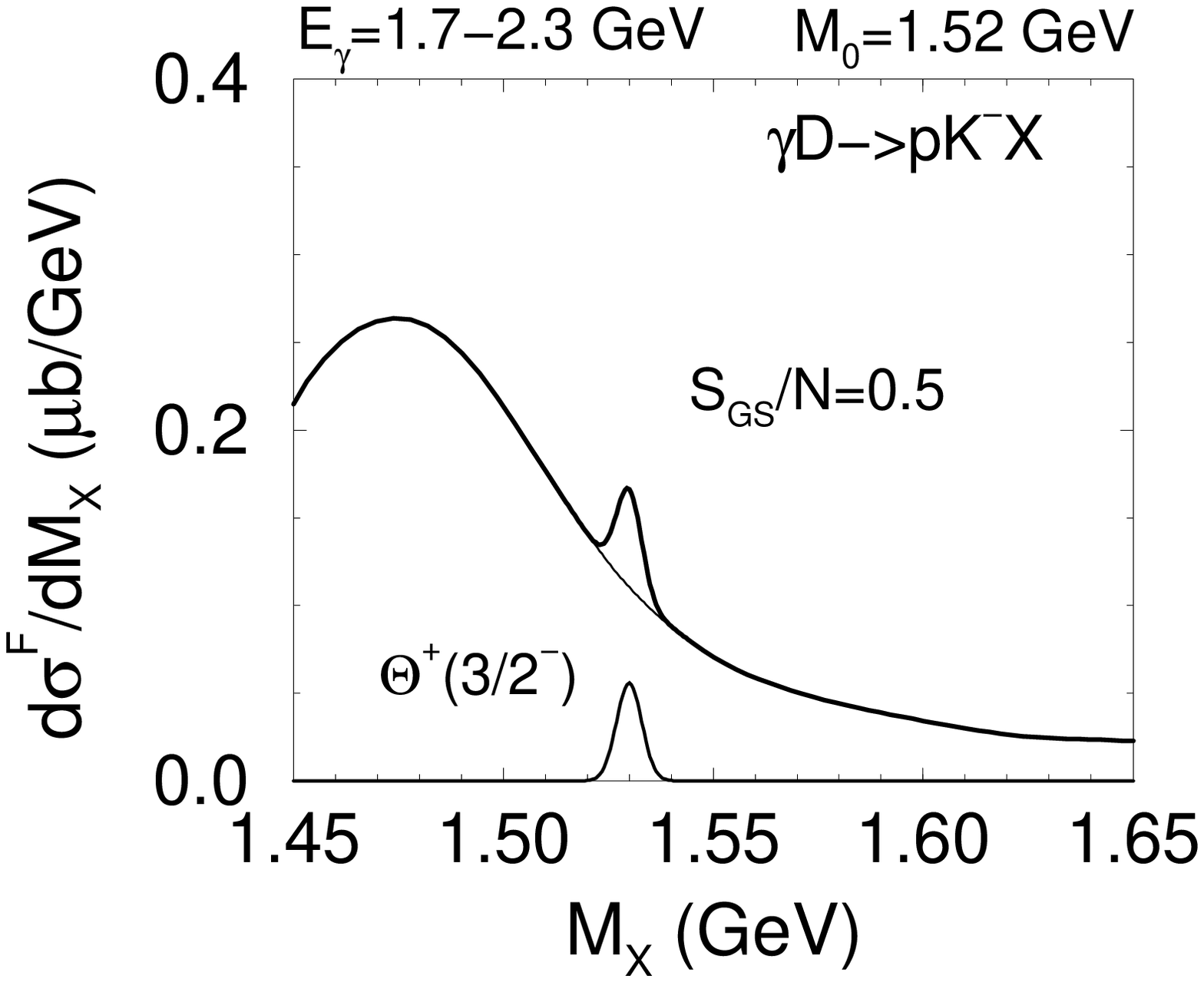}\qquad
  \includegraphics[width=0.4\columnwidth]{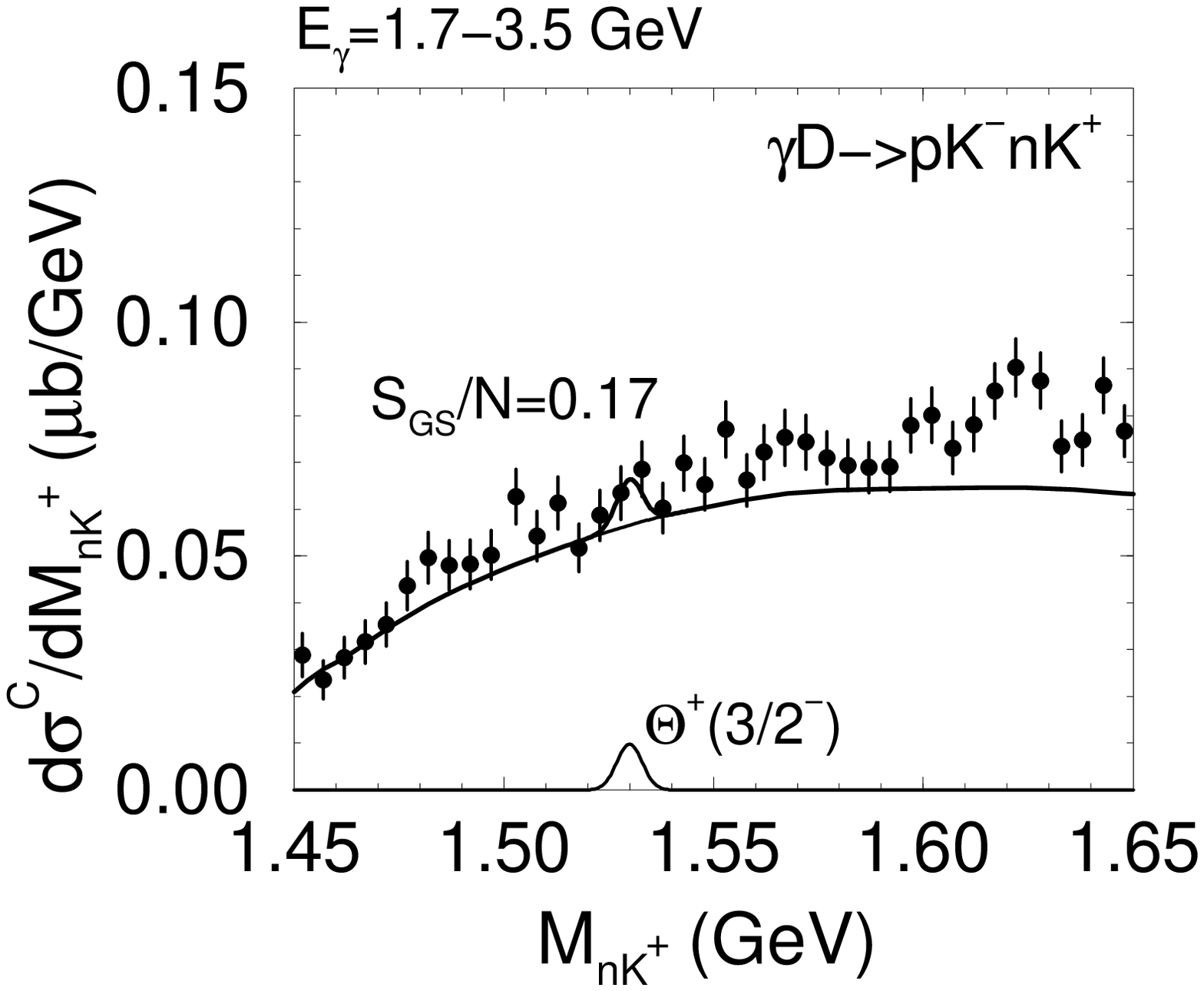}
  \caption{\label{FIG:26}{\small%\tcaps%
   Left panel:
   Missing mass distribution in inclusive $\gamma D\to  pK^-X$
   at $E_\gamma=1.7-2.3$~GeV  and
   the $pK^-$ photoproduction angular cut
  ($\theta_{pK^-}<22^0$~(c.m.s.))   and
   $\phi$-meson cut.
   Right panel: $nK^+$ invariant mass
   distribution in exclusive $\gamma D\to  pK^-nK^+$
   at $E_\gamma=1.7-3.5$~GeV and for CLAS experimental
   conditions(i)-(v).
   Experimental data from
   Ref.~\protect\cite{JLab-06}.
   In both cases, $J_{\Theta}^P=3/2^-$ and
   the $\Tp$ signal is folded with a Gaussian
   resolution function with a width of 3~MeV.}}}
 \end{figure}
 The $\Tp$ photoproduction cross section at maximum is equal to
\begin{equation}\label{Theta-signalJL}
  \frac{d\sigma^{\Tp\,C}}{dM}{\Big|_{\rm max}}\simeq 29\,
  \frac{{\rm n b}}{\rm GeV}~.
\end{equation}
Using this value we can evaluate the total $\Tp$ formation cross
section for conditions (i)-(v)
 \begin{eqnarray}
 \sigma^{\Tp\,C}_{\rm tot}
 &\simeq& 2\times\frac{\pi}{2}\times\Gamma_\Tp\times 29\,\,
 \frac{{\rm nb}}{\rm GeV}
 \simeq~0.1~{\rm nb}~.
\label{CS-CLAS}
 \end{eqnarray}
 In the latter case, the additional factor 2 means that only the
 $\Tp\to nK^+$ decay channel is under consideration.
 Our estimate of the $\Tp$ formation cross section
 for the CLAS experiment is three times smaller than the upper
 bound for the $\Tp$ signal (0.3~nb) reported in~\cite{JLab-06}.

 For illustration, in Fig.~\ref{FIG:26} we exhibit simultaneously the
missing mass distribution in the inclusive $\gamma D\to pK^-X$
reaction, averaged over the interval $E_\gamma=1.7-2.3$~GeV with
the $pK^-$ photoproduction angular cut
($\theta_{pK^-}<22^0$~(c.m.s.)) (left panel) and the $nK^+$
invariant mass in the exclusive $\gamma D\to  pK^-NK^+$ reaction
for the CLAS experimental conditions (i)-(v)  (right panel)
together with the available experimental data \cite{JLab-06}.
Remember, that the $nK^+$ invariant mass distribution shown
in~\cite{JLab-06} is obtained after removing certain processes:
contributions from $\phi$ meson and $\Ls$ excitations, and the
neutron spectator channels, because this experiment intends to
search for direct $\Tp$ production in the $\gamma n\to \Tp K^-$
reaction. In our analysis of the $\Tp$ formation process in the
CLAS experiment we include all experimental conditions (i)-(v). In
order to be close to the conditions of the data analysis in
Ref.~\cite{JLab-06} we include in the consideration the acceptance
correction  factor which restores the full 4-body phase space
broken by the cuts (i)-(v). Our results with acceptance
corrections are shown in Fig.~26. \footnote{The acceptance
correction increases
 the total cross section in Eq.~(\ref{CS-CLAS}) to 0.15~nb.}.
 The model satisfactory describes the data at low $M_{nK^+}$ and
slightly underestimates them at higher invariant mass with
$M_{nK^+}>1.6$~GeV. It is not surprising because our simple model
does not pretend for detailed description of all aspects of the
$\gamma D$ interaction in a wide interval of the photon energy.
Our main purpose is the analysis of $\Tp$ signal-to-noise ratio at
$M_{nK^+}\sim M_\Theta$, where  the model looks quite reasonable.
% Some deviation between calculation and data may be
%explained by difference in acceptance corrections. Note however,
%that since the acceptance corrections in data analysis don't
%distinguish between the background and $\Tp$ signal channels then
%the signal-to-noise ratio which is the main purpose of our study,
%does not depend on the particular method (model) of this
%correction, while the absolute value of the cross section does
%\footnote{The acceptance correction increases the total cross
%section in Eq.~(\ref{CS-CLAS}) to 0.15~nb.}.

 For the left and right panels of Fig.~26
 the $\Tp$ signal is folded with a Gaussian resolution function
with $\sigma=3$~MeV. Notice, that utilizing the Gaussian
resolution function reduces the $\Tp$ signal at the maximum
position by a factor $0.21$ (cf. Eq.~(\ref{RES13}))\footnote{Our
choice $\sigma=3$~MeV is rather illustrative. Using
Eq.~(\ref{RES13}) one can easily re-estimate the amplitude of the
$\Tp$ signal for any value of $\sigma$.}. From
Fig.~\ref{FIG:26}~(right) one can see that the effective $\Tp$
signal predicted for the CLAS conditions (i)-(v) is comparable to
the statistical fluctuations.
%In
%principle, few data points around $M_{pK^-}\simeq1.53$~GeV
%visually follow the resonance signal. But it is clear that this
%``signal'' can not be considered seriously, because it is
%comparable to (or even smaller than) the statistical fluctuations
%due to available experimental acceptance above and below the $\Tp$
%resonance position.
Therefore,  absence of a bright $\Tp$ signal in the CLAS
data~\cite{JLab-06} does not exclude its possible manifestation
under more favorable experimental conditions.

\section{Summary}

In summary, we analyzed the possible manifestation of the $\Tp$
formation process in inclusive $\gamma D\to pK^-X$ reactions. If
the $\Tp$ exists, then in the $[\gamma D,pK^-] $ missing mass
distribution there must be a distinct $\Tp$ peak. Its strength
depends on the $\Tp$ spin and parity, and has a maximum value for
$J_\Theta^P=3/2^-$. We found that at forward angles of the $pK^-$
pair photoproduction the signal-to-noise ratio  is most favorable.

We also analyzed the recent results of the CLAS collaboration and
found that the present experimental conditions are not favorable
for  studying the $\Tp$ \underline{formation} processes. The
corresponding signal to noise ratio is small, and the $\Tp$ signal
is comparable to the statistical fluctuations due to the
experimental acceptance.

In our model a distinct $\Tp$ signal is caused by the constructive
interference of the $\Ls$ photoproduction from the proton and
neutron in the associated $\Ls\Tp$ photoproduction off the
deuteron. In  calculations we use relatively old data for $\Ls$
photoproduction off the proton at photon energies greater than the
most favorable ones for the associated $\Ls\Tp$ photoproduction
making a corresponding extrapolation. For a more detailed study of
this effect,  new high-statistics low-energy data both for $\gamma
p\to\Ls K^+$ and $\gamma n\to\Ls K^0$  are greatly desired,
especially for large kaon photoproduction angles. A similar
problem concerns fixing the non-resonant background. The
elementary $\gamma p\to p K^+K^-$ cross section is very important
here. In our analysis we used old data with low accuracy. It is
clear that for understanding the $\Tp$ formation processes one
needs more accurate low-energy data for this elementary
subprocess, too. However, our main results have a general
character. Thus, it seems more reliable to detect the $\Tp$ signal
in the $KN \to \Tp$ fusion reaction realized in associated
$\Ls\Tp$ photoproduction, which may be seen in inclusive $\gamma
D\to pK^-X$ reaction for certain experimental conditions.

 Finally, we note that the $\Tp$ formation
 reaction together with other accompanying processes
 considered in the present paper  may be studied experimentally
 at the electron and photon facilities at LEPS of SPring-8, JLab,
 Crystal-Barrel of ELSA, and GRAAL of ESRF.
% European Synchrotron Radiation Facility

\acknowledgments

 We appreciate many fruitful discussions with T.~Nakano,
 and we  thank D.~Diakonov, H.~Ejiri,
 M.~Fujiwara, K.~Hicks, A.~Hosaka, T.~Mibe, R.~Mutou, M.~Naruki,
 T.~Sato, and K.~Yazaki
 for useful comments. We appreciate
 B.~Mecking for careful reading our manuscript and valuable
 suggestions.
 One of the authors (A.I.T.) thanks H.~En'yo for
 offering the hospitality at RIKEN.
 This work was supported by BMBF grant 06DR121, GSI-FE.

\appendix

\section{Kinematics for the reaction $\bm\gamma D\to pK^-NK$ }

Let us consider the determination of the momenta of all outgoing
particles in the $\gamma D\to pK^-NK$ reaction at fixed input
parameters defined in Sec.~II in  detail.  The square of the total
energy in the c.m.s.\ is
\begin{equation}\label{E3}
  s_D=M_D^2 + 2M_DE_\gamma~,
\end{equation}
where $M_D$ is the deuteron mass and $E_\gamma$ is the photon
energy in the laboratory system. The momenta $p_i$ and $p_f$ read
\begin{eqnarray}\label{E4}
  {p}_i&=&\frac{\sqrt{\lambda(s_D,M_D^2,0)}}{2\sqrt{s_D}}
  =\frac{s_D-M_D^2}{2\sqrt{s_D}},\nonumber\\
  {p}_f&=&\frac{\sqrt{\lambda(s_D,M_X^2,M_Y^2)}}{2\sqrt{s_D}},
\end{eqnarray}
 where $\lambda(x^2,y^2,z^2)=(x^2-(y-z)^2)(x^2-(y+z)^2)$
 is the triangle kinematical function.
 The four momenta of proton and $K^-$ meson in the $Y$ rest frame
 is defined by the mass $M_Y$ and the solid angle $\Omega_Y$.
 Thus, the  absolute value of
 the decay 3-momentum reads
 \begin{equation}\label{E5}
 \widetilde{\bar q}=
 \frac{\sqrt{\lambda(M_Y^2,M_K^2,M_N^2)}}{2M_Y}.
 \end{equation}
 Then the four momenta are defined as
 \begin{eqnarray}\label{E6}
 \widetilde{p}_{K^-}&=&( \widetilde{E}_{K^-},\,\,
 \widetilde{\bar q}\sin\theta_Y\cos\varphi_Y,
 \widetilde{\bar q}\sin\theta_Y\sin\varphi_Y,
 \widetilde{\bar q}\cos\theta_Y),\nonumber\\
 \widetilde{E}_{K^-}&=&\sqrt{\widetilde{\bar q}^2 +M_K^2 }
\end{eqnarray}
and
\begin{eqnarray}\label{E7}
 \widetilde{p_{p}}&=&( \widetilde{E_{p}},\,\,
 -\widetilde{\bar q}\sin\theta_Y\cos\varphi_Y,
 -\widetilde{\bar q}\sin\theta_Y\sin\varphi_Y,
 -\widetilde{\bar q}\cos\theta_Y),\nonumber\\
 \widetilde{E_{p}}&=&\sqrt{\widetilde{\bar q}^2 +M_N^2 },
\end{eqnarray}
 respectively. Note, that here the $\widetilde {\bf z}$ axis is taken
 along ${\bf p}_Y$, and the $\widetilde{\bf y}$ axis coincides with ${\bf
 y}$.
 Next, we boost these momenta to the c.m.s. along the ${\widetilde {\bf z}}$ axis
 as
\begin{eqnarray}\label{E8}
  {p_p}'_0&=&\gamma_Y(\widetilde{p_p}_0 + v_Y \widetilde{p_p}_3),\qquad
  {p_p}'_3=\gamma_Y(\widetilde{p_p}_3 + v_Y \widetilde{p_p}_0),\nonumber\\
  {p_p}'_1&=&\widetilde{p_p}_1\qquad
  {p_p}'_2=\widetilde{p_p}_2,
\end{eqnarray}
 where $v_X=p_f/\sqrt{p_f + M_X^2}$ and
 $\gamma_X=v_X/\sqrt{1-v_X^2}$. Then, we rotate the coordinate
 system around the ${\bf y}$ axis by the angle $\theta$ to get the
 momenta in the c.m.s. with ${\bf z}$ along the photon momentum $\bf k$
\begin{eqnarray}\label{E9}
 {p_p}_1&=&{p_p}'_1\cos\theta + {p_p}'_3\sin\theta,\nonumber\\
 {p_p}_3&=&{p_p}'_3\cos\theta - {p_p}'_1\sin\theta,\nonumber\\
 {p_p}_2&=&{p_p}'_2,\qquad {p_p}_0={p_p}'_0.
\end{eqnarray}
 Similarly, we transform the momenta of the outgoing nucleon $N$ and
 $K$ meson of the $X$ system with obvious substitutions:
 $M_Y\to M_X$,
 $\theta_Y\to \theta_X$,
 $\varphi_Y\to \varphi_X$,
 and $\theta\to \pi +\theta$.

 %The transformation from c.m.s. to the laboratory frame is standard.

\section{Amplitudes of the elastic scattering\\ processes}

\subsection{$\bm {K^-N\to K^-N}$ scattering}

Let us consider the elastic $K^-p\to K^-p$ scattering. The
amplitude is related to the differential cross section via
\begin{equation}\label{C1}
  \frac{d\sigma^{Kp}}{d\Omega}=
  \frac{1}{64\pi^2\,s}\frac{{ p}_f}{{ p}_i}
  \frac12\sum\limits_{m_im_f}
  |T_{m_fm_i}^{Kp}(s,\cos\theta)|^2
\end{equation}
where ${ p}_i,m_i$ and ${ p}_f,m_f$ are the three dimensional
relative momenta and the proton spin projections in   the initial
and the final states, respectively, $s$ is the  square of the
total energy in the c.m.s., and $\theta$ denotes the scattering
angle. In our calculations we take the differential cross section
from experiment \cite{Kp_scattering}, given as a function of the
scattering angle in certain energy intervals. In rescattering
processes, one of the incoming proton may be off-shell, and
therefore, generally, ${ p}_i\neq { p}_f$. Thus, the scattering
angle is defined as
\begin{eqnarray}\label{C2}
  \cos\theta = \frac{2E_iE_f -2 M_K^2+ t}{{ p}_i{  p}_f}
\end{eqnarray}
 with $t$, ${ p}_{i(f)}$  and $E_{i(f)}$ given as
 \begin{eqnarray}\label{C3}
 t&=&(q-q')^2~,\nonumber\\
 |{p}_i|&=&\frac{\sqrt{\lambda(s,M_K^2,{\bar {M}_N}^2)}}
 {2\sqrt{s}}~,\nonumber\\
 |{ p}_f|&=&\frac{\sqrt{\lambda(s,M_K^2,M_N^2)}}
 {2\sqrt{s}}~,\nonumber\\
 E_{i(f)}&=& \sqrt{{\bf p}_{i(f)}^2 + M_K^2}~,
 \end{eqnarray}
where $q$ and $q'$ are the kaon four momenta in initial and final
states, respectively, and $\bar {M}_N^2$ is the square of the four
momentum of the incoming off-shell nucleon.

We use the following parameterization of the differential cross
section (in  mb)
\begin{eqnarray}
 p_L<0.3663~,\qquad\qquad
 {d\sigma^{Kp}}/{d\Omega}&=&
 {3.01}(2 +\cos\theta)/s~,\nonumber\\
 0.3663<p_L<0.4185~,\qquad
 {d\sigma^{Kp}}/{d\Omega}&=&
 {3.01}(1 +3\cos^2\theta)/s~,\nonumber\\
  0.4185<p_L~,\qquad\qquad
 {d\sigma^{Kp}}/{d\Omega}&=&
 0.5(1 +(1+\cos\theta)^\delta)~,\label{C4}
\end{eqnarray}
where $p_L$ is the kaon momentum in laboratory frame in GeV/c and
\begin{equation}\label{C5}
\delta=1.443\ln(13.33p_L-1)~.
\end{equation}
The spin dependence of the amplitude is chosen in the simplest form as
\begin{equation}\label{C6}
  T_{m_fm_i}^{Kp}=T_0\,
  (\delta_{m_i,m_f} + \delta_{-m_i,m_f})~.
\end{equation}

Concerning the $K^-n$ reaction, we employ
$d\sigma^{K^-n \to K^-n}\simeq d\sigma^{K^-p\to K^-p}$.

\subsection{${\bm pN\to pN}$ scattering}

For the differential cross section of the elastic $pp$ scattering
we use the parameterization of Ref.~\cite{pp_scattering}:
\begin{equation}\label{C7}
  \frac{d\sigma}{dt}=A\exp(Bt_a + Ct_a^2)~,
\end{equation}
where $t_a=2M_N^2 - 2E^2 + 2p^2|\cos\theta|$, $p$ and $E$ are the
proton  momentum and energy in the c.m.s., respectively. The
parameters $A,B$, and $C$ are listed in Table.~I.
\begin{table}[t]
  \caption{Parameters of the function ${d\sigma}/{dt}=A{\rm e}^{Bt_a + Ct_a^2}$ .}
\begin{tabular}{|c|c|c|c|}
  \hline\hline
  $p_L$ & A & B & C \\
  (MeV/c) & $[$mb/(GeV/c)$^{2}]$ & (GeV/c)$^{-2}$ & (GeV/c)$^{-4}$
  \\ \hline
  $<$899      & 87.1  & 0.96 & 2.91 \\
  $900-  999$ & 78.8  & 1.17 & 1.89 \\
  $1000-1099$ & 64.7  & 1.00 & 0.72 \\
  $1100-1199$ & 60.4  & 2.01 & 2.10 \\
  $1200-1299$ & 79.6  & 4.38 & 3.96 \\
  $1300-1399$ & 107.4 & 5.76 & 3.68 \\
  $1400-1499$ & 116.1 & 5.45 & 2.04 \\
  $1500-1599$ & 126.2 & 6.51 & 3.56 \\
  $1600-1699$ & 137.0 & 7.20 & 3.80 \\
  $1700-1799  $ & 140.9 & 7.66 & 4.39 \\
  $1800-2399  $ & 141 & 7.66 & $3.55$ \\\hline
\end{tabular}
%  \caption{Parameters of function in Eq.~(\protect\ref{C7}).}
\end{table}
Here $p_L$ is the proton momentum in laboratory system.

For $p_L=2.2-3.5$~GeV/c the cross section is parameterized as a
sum of two exponentials
\begin{equation}\label{C8}
  \frac{d\sigma}{dt}=A[(1-\alpha)\exp(Bt_a) + \alpha\exp(Ct_a)]~,
\end{equation}
where $A\simeq141$~mb/GeV$^2$ and $\alpha$, $B$, and $C$ are
listed in Table~2.
\begin{table}[t]
  \caption{Parameters of the function ${d\sigma}/{dt}=
  A[(1-\alpha)\,{\rm e}^{Bt_a} + \alpha\,{\rm e}^{Ct_a}]$ .}
\begin{tabular}{|c|c|c|c|}
  \hline\hline
  $p_L$ & $\alpha$ & B & C \\
  (MeV/c) &   & (GeV/c)$^{-2}$ & (GeV/c)$^{-2}$
  \\ \hline
  $2200-2599  $ & 0.022 & 7.8 & 0.7 \\
  $2600-2999  $ & 0.015 & 8.0 & 0.7 \\
  $3000-3500  $ & 0.015 & 8.8 & $1.0$ \\\hline
\end{tabular}
%  \caption{Parameters of function in Eq.~(\protect\ref{C7}).}
\end{table}

Similarly to the $K^-p$ scattering we ignore the spin dependence
of the amplitude. In our study we assume approximately
$d\sigma^{pn}\simeq \,d\sigma^{pp}$.

\end{document}